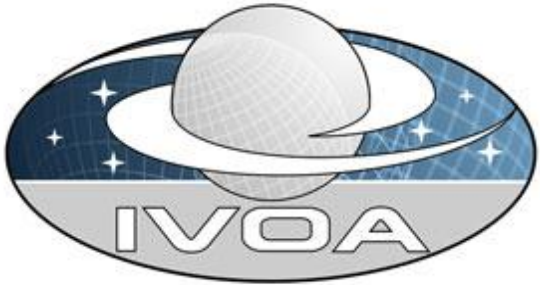

**I**nternational

**V**irtual

**O**bservatory

**A**lliance

# Simulation Data Model

# Version 1.00

*IVOA Recommendation 2012 May 3$^{rd}$*

**This version:**
http://www.ivoa.net/Documents/SimDM/

**Latest version:**
http://www.ivoa.net/Documents/SimDM/20120503/

**Previous version(s):**
http://www.ivoa.net/Documents/SimDM/20120302/
http://www.ivoa.net/Documents/SimDM/20111019/
http://www.ivoa.net/Documents/SimDM/20110909/
http://www.ivoa.net/Documents/SimDM/20110428/
http://www.ivoa.net/Documents/SimDM/20110427/

**Working Group**:
   http://www.ivoa.net/cgi-bin/twiki/bin/view/IVOA/IvoaDataModel
   http://www.ivoa.net/cgi-bin/twiki/bin/view/IVOA/IvoaTheory

**Editors:**
   Gerard Lemson, Hervé Wozniak

**Authors:**
   Gerard Lemson, Laurent Bourgès, Miguel Cerviño, Claudio Gheller, Norman Gray, Franck LePetit, Mireille Louys, Benjamin Ooghe, Rick Wagner, Hervé Wozniak

# Abstract

In this document and the accompanying documents we describe a data model (Simulation Data Model) describing numerical computer simulations of astrophysical systems. The primary goal of this standard is to support discovery of simulations by describing those aspects of them that scientists might wish to query on, i.e. it is a model for *meta*-data describing simulations.
This document does *not* propose a protocol for using this model. IVOA protocols are being developed and are supposed to use the model, either in its original form or in a form derived from the model proposed here, but more suited to the particular protocol.

The SimDM has been developed in the IVOA Theory Interest Group with assistance of representatives of relevant working groups, in particular DM and Semantics.

# Link to IVOA Architecture

The figure below shows where SimDM fits within the IVOA architecture:

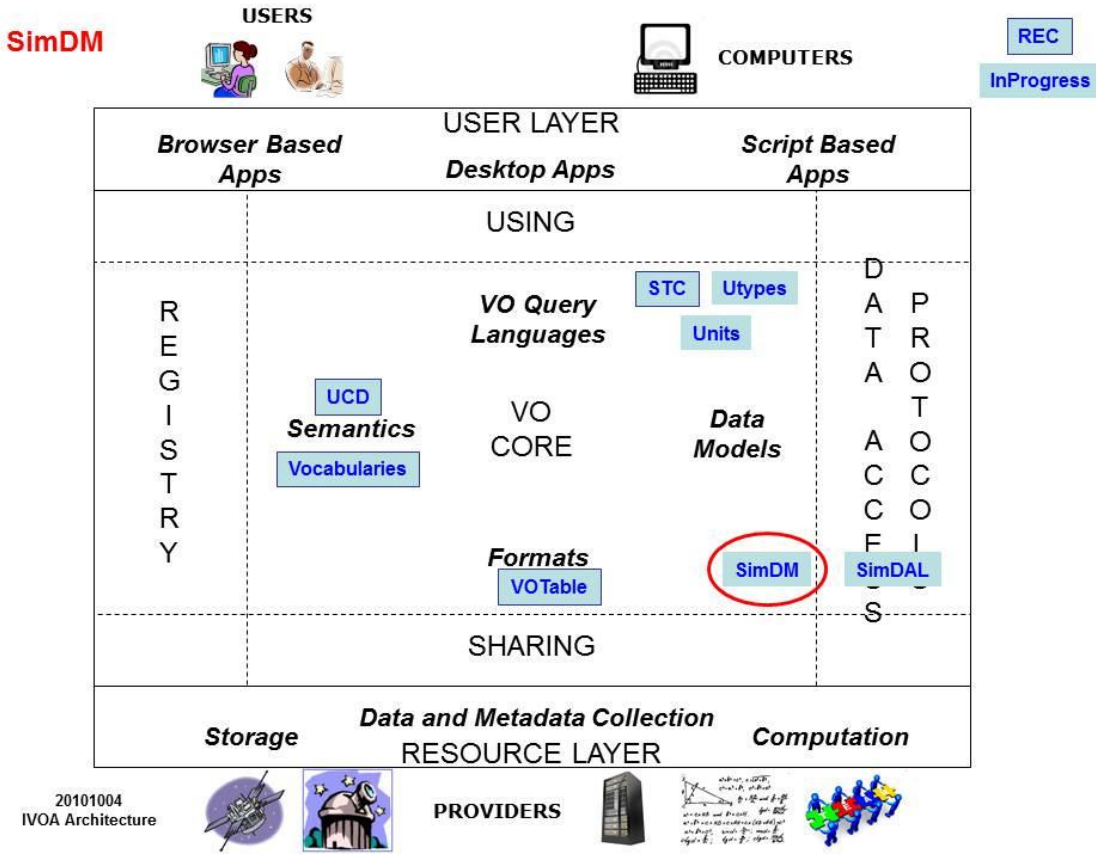

The data model proposed here is intended to be used in some IVOA standard services such as:

- The "Simulation Database (SimDB)" protocol, which describes a particular web service that gives access to a database containing metadata describing simulations. SimDB will be a specification for an online web service providing access to a repository storing metadata about numerical computer simulations of astrophysical systems and related resources. A SimDB is supposed to be used to discover simulations together with web services providing access to them. Therefore, it is located in the Registry part of the IVOA Architecture.
- Any Data Access Layer (DAL) protocol dedicated to theoretical products, currently joined under the common name SimDAL. The more detailed specification of these services, fully compliant with the current approach promoted by the DAL Working Group, is the goal of the SimDAL specification.

## Status of This Document

This is an IVOA Recommendation reviewed by IVOA Members. It has been endorsed by the IVOA Executive Committee as an IVOA Recommendation. It is a stable document and may be used as reference material or cited as a normative reference from another document. IVOA's role in making the Recommendation is to draw attention to the specification and to promote its widespread deployment. This enhances the functionality and interoperability inside the Astronomical Community.

The first release of this document was 2011 April 28.

A list of current IVOA Recommendations and other technical documents can be found at http://www.ivoa.net/Documents/ .

## Acknowledgements

We thank various persons for useful discussions in the course of this work: first, the participants of the Feb 2006 theory workshop in Cambridge, UK, where this work was started; second, the participants of the April 2007 SNAP workshop in Garching, Germany, where the design started taking shape. The work has also been influenced by the participants of the Technical Coordination Group of the EuroVO-DCA project and participants of the theory workshop organised in the context of that project in Garching, 2008. Then we want to thank particularly the following persons for useful discussions and feedback: Jeremy Blaizot, Miguel Cerviño, Klaus Dolag, Pierro Madau, Adi Nusser, Ray Plante, Volker Springel, and Alex Szalay. We finally want to thank participants to the theory sessions in all the interoperability meetings since Victoria 2006, where parts of this work were discussed.

## Conformance related definitions

The words "MUST", "SHALL", "SHOULD", "MAY", "RECOMMENDED", and "OPTIONAL" (in upper or lower case) used in this document are to be interpreted as described in IETF standard, RFC 2119 [38].

The **Virtual Observatory (VO)** is a general term for a collection of federated resources that can be used to conduct astronomical research, education, and outreach. The **International Virtual Observatory Alliance (IVOA)** is a global collaboration of separately funded projects to develop standards and infrastructure that enable VO applications. The International Virtual Observatory (IVO) application is an application that takes advantage of IVOA standards and infrastructure to provide some VO service.

# Contents

# 1 Introduction

In this document we describe an IVOA standard data model for describing simulations[1]. The primary goal of this standard is to support discovery of simulations by describing those aspects of them that scientists might wish to query on, i.e. it is a model for *meta*-data describing simulations. This document does *not* propose a protocol for using this model. IVOA protocols are being developed and are supposed to use the model, either in its original form or in a form derived from the model proposed here, but more suited to the particular protocol.

The other documents related to this proposal, and being a part of the specification, are given in Table 1. They can be found at the root URL http://ivoa.net/Documents/SimDM/20120503/.

Additionally, an Implementation Note explains how to use this model to describe various kinds of theoretical products [7].

Section 2 described our methodology whereas the model itself is detailed in Section 3. A few specific issues of serialization are addressed in Section 4. The development of SimDM is linked to other IVOA efforts that deserve to be mentioned. Section 5 deals with that point. In the Appendices we deal with the scientific motivation at the origin of creating SimDM and the 4-years history of the developments (Appendix A), more details on the UML profile (Appendix B) and some issues (Appendix C).

---

[1] We will use the term *simulations* for the running of a simulation code as well as for their results. And we will often include post-processing codes and their results as well.

**Table 1: list of documents as part of the SimDM specification, accessible from http://ivoa.net/Documents/SimDM/20120503/**

| | | |
|---|---|---|
| SimDM.html | Full browsable specification of the model | html/SimDM.html |
| SimDM_DM.png | Graphic view of the whole model (large image) | uml/SimDM_DM.png |
| SimDM_DM.xml | MagicDraw UML diagram serialised to XMI | uml/SimDM_DM.xml |
| SimDM_INTERMEDIATE.xml | Intermediate representation of the model: a (generated) XML document representing the complete model in more readable format than XMI | uml/SimDM_INTERMEDIATE.xml |
| intermediateModel.xsd | XML schema document for intermediate representation's XML format | uml/intermediateModel.xsd |
| xsd/ | XML schema documents (generated) representing mapping of UML to XSD | xsd/ |

# 2 SimDM: application, approach and outline

## 2.1 Phase 1: analysis

The analysis phase investigates the "world the application lives in", its "universe of discourse" [28] and describes it in a domain model. To get constraints on this universe and its contents we follow [36] in trying to gather some 20 science questions that the application should be able to answer. The application is here a system consisting of the data model together with the protocol and implementations. The model will be designed in such a way that it can contain the required information. The protocol and implementations must support efficient querying for this information. [36] used this approach in the design of the SDSS database.

To create such a list of questions we have contacted scientists with the question that if they were presented with a database of simulation metadata, what questions they would want to ask of it to find interesting simulations. The following list summarises their answer:

- What system/object is being simulated?
- What physical processes are included?
- How is the system being represented in the simulation (particles (Lagrangian), (adaptive) mesh (Eulerian)), both, other?
- How are the physical processes implemented?
- What numerical approximations were used (e.g. resolution, softening parameter)?
- What observables are available for the system/object, possibly as function of time[2]? As it is a spatial system, at least *simulation box*size, centre-of-mass position.
- What observables are available for the constituents, i.e. what is the schema of the objects from which the simulation built e.g. particles in N-body simulation, grid cells in an adaptive mesh simulation or particle groups in a cluster finder?
- Per snapshot, per simulation object type, per variable:
    - Characterise the *possible* values
    - Characterise the result
- Are post-processing results available?
- Are services/applications available for accessing the results?
- Which code ran the simulation?
    - Which *version* of the code?
    - Is software available?

---

[2] Re: Rick Wagner's example of certain properties only being calculated after a certain stage in the simulation is reached.

- Who ran the simulations?
- What were values of input parameters?
- How were initial conditions created?
- How the results are parameterized?
- Can I access grids of models? Can I access individual results?
- Which are the inputs ingredients (usually, which data collections are used?)
- How I can run a simulation? Can I do it on-the-fly?
- Can I include my simulations in the VO in an easy way? What I should do?
- Can I compare different simulations? Can I compare the simulation with my data?
- Which simulations provide diagnostic tools? (i.e. distance/extinction/quasi-scale free quantities)
- Can I combine the results of different simulations in a single file adapted for my needs (e.g. own code)?

## 2.2 Phase 2: domain model

The result of the analysis phase is a model in its own right, albeit rather sparse and schematic. For this purpose we have built on previous work by adapting the so called *Domain model for Astronomy* proposed in [12]. This model forms the basic structure of the domain model for SimDM, illustrated in Figure 1.

Figure 1 is used in a narrative motivating the final structure of the full SimDM. We start by assuming the existence of one or more **Files** that a publisher thinks may be of interest to the community because they contain astronomical data. Instead of in files the data might also reside in a **Database**, and to be generic we introduce a **Storage** base class that abstracts the actual physical location of the data.

Registering that files exist somewhere is not of great interest without providing information about the *contents* of the files. The philosophy that we follow is that the files are of potential interest because they contain the **Results**[3] of an (astronomical) **Experiment**, and accordingly their contents must be explained by describing the experiment that gave rise to it. Only in this way can one make scientific use of the files or other storage resources.

The abstract **Experiment** is made concrete by adding some examples of experiment types that are important for the current model dealing with **Simulations** and simulation **PostProcessing**.

In our model, **Experiment** represents the actual *running* of an experiment; to describe the *design* of the experiment (the so-called *experimental protocol*) we introduce the concept of *(experimental)* **Protocol**[4]. This separation between design of experiment and the execution is a normalisation that reduces redundancy in the model. We mirror the concrete subclasses of **Experiment** by adding concrete subclasses to *(experimental)* **Protocol** such as **Simulator**, which represents simulation codes according to which **Simulations** are run, and **PostProcessor** corresponding to **PostProcessing** runs.

---

[3] We do not assume that in reality the relation between the conceptual Result and the concrete Storage elements can be modelled by a single reference. Especially for the largely non-standardised world of simulations a single result can be distributed over many files, but it is also possible for one file to contain multiple results. In the current SimDB model we do not attempt to model such relations explicitly. We delegate the responsibility for accessing the physical results to (web) services and this issue is more explicitly addressed by a DAL protocol.

The *(experimental)* **Protocol** class contains **InputParameters**. An **Experiment** using a particular *(experimental)* **Protocol** only needs to indicate the *values* for these parameters. In this way a single instance of the *(experimental)* **Protocol** can be reused by many **Experiments** performed according to it.

The *(experimental)* **Protocol** also defines the possible structure of the results of the experiments. In our model **Results** contain one or more **ObjectCollection**-s containing **Objects** of a given type, represented by the **ObjectType** contained by (*experimental)* **Protocol**. The **ObjectType** defines the **Properties** that these objects have. The **Object** finally can assign values to these Properties using the **ValueAssignment**.

For example the results of N-body simulations may contain particles having properties position, velocity, mass and possibly others. Adaptive Mesh Refinement (AMR) simulations produce results that are collections of mesh cells of various sizes, positions and contents. Similarly post-processing codes such as halo finders produce "halos" and "semi-analytical" galaxy formation codes produce galaxies.

In general a single result can contain objects of different types. For example a Smooth Particle Hydrodynamics (SPH) simulation may contain dark matter particles, star particles and gas particles. And in general the codes allow one to configure which of these exactly are chosen in a given experiment.

One aspect of the experiment that is not determined by the experimental protocol is *why* the experiment was performed. In the model we introduce the **Target** concept for this, which represents real world objects or processes that are being simulated. For example, with the same N-body simulator one may simulate a galaxy merger or the evolution of large scale structure of the universe.

As discussed above, the actual way in which results are stored in files or databases is hard, if not impossible to model. Instead we assume that **Webservices** of various kinds may be used to access the results of simulations.

Some of these will be standardised in any DAL specification, but custom services may also be introduced. The model allows one to describe the experiments and their results, which should allow users to discover results of interest, after which the web services can be called for actually accessing these.

---

[4] Further on, the word *protocol* will be preceded by the adjective *experimental* (in parenthesis and italicized) to keep clear the distinction with any other IVOA protocol.

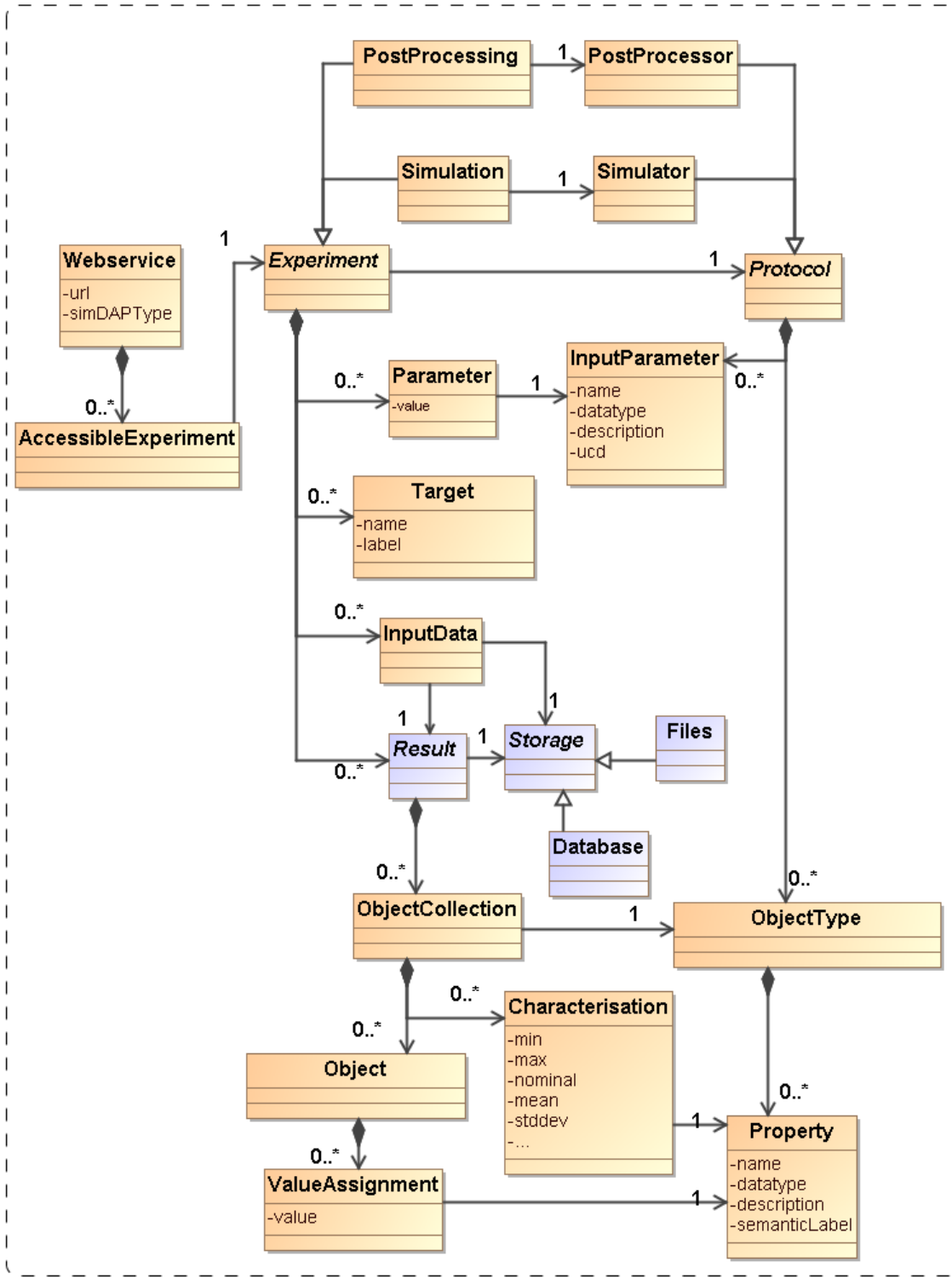

**Figure 1:** Schematic domain model encapsulating the main design constructs in SimDM. Elements coloured orange are represented directly in SimDM, possibly with a different name. Purple elements are not part of that model, but are used to explain and motivate other features that do appear there.

# 3  Logical model overview

SimDM, is a logical model in the sense of [34] and based on the domain model described in 2.2. SimDM is "logical" in that it aimed to support an application, namely SimDB, a repository of simulation metadata, but is still implementation neutral and represented in UML. As a model for an implementation it is fully detailed. It has a human readable HTML representation which contains the detailed description of all elements [5]. That document should be consulted for the details of the model.

Here we introduce the main concepts and motivate the main design decisions. Where possible we try to add a hyperlink from a concept's name pointing into the HTML document the first time we use the name. The link will consist of a root URL to the location of the HTML document, followed by a #<UTYPE> that identifies the description of the actual concept in the HTML document. This we feel is very much in the spirit of the use cases of UTYPEs. Later references to the concepts will in general not contain the link. Then class names will be capitalised. Abstract classes will be in italics. Names of packages, attributes, references or collections will be preceded by the class name where necessary or it will be assumed to be clear from the concept what is intended.

For illustration and examples we use UML instance diagrams rather than XML. See Appendix B.14 for an explanation of this type of diagram. For reasons explained in section 4.2, serialising this model to XML is rather complex and would likely confuse the reader rather than elucidate her upon first reading. XML serialisations can be found in an implementation Note [7] that is produced separately.

## 3.1  Packages

UML Packages are subsets of classes and data types that are deemed to belong together. Whilst not essential to the model, we have used them to provide some level of modularity. Their main role is played in the XML schemas derived from the model. Each package has its own type-schema (see 4.2) which provides a somewhat finer level of reuse.

The diagram in Figure 2 shows the packages we use and their dependencies. This hierarchy is reflected in the UTYPEs, see section 4.1. The colours assigned to the packages correspond to the colours of classes in the diagrams in later sections. The subdivision in the one parent and three child packages follows the resource *class* hierarchy described next.

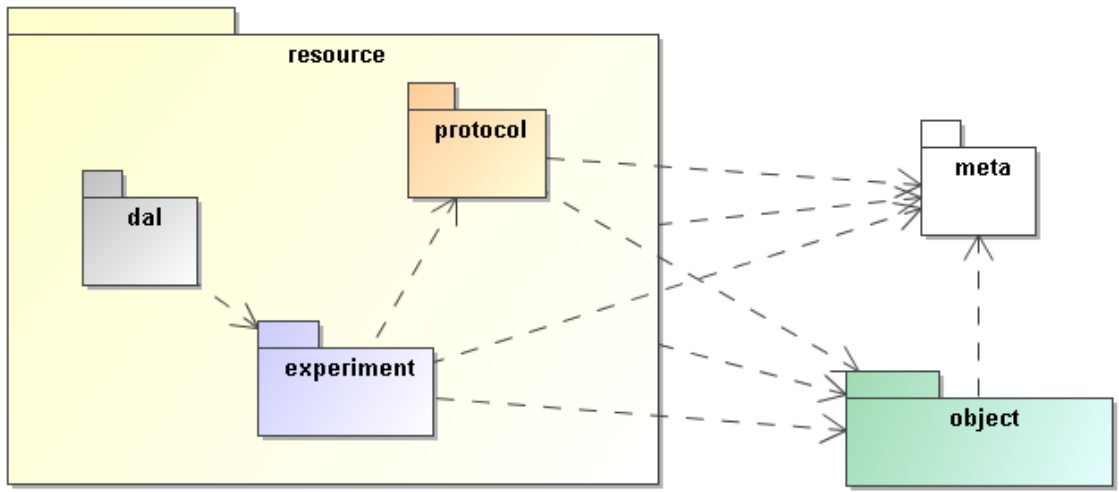

**Figure 2: The packages of the SimDM and their relationships. These are related to each other through directed dependency links indicated by the dashed arrows.**

## 3.2 Resource

The SimDM aims at describing simulations and related concepts. The current model does so with of the order of 40 separate object types, or classes. Most of these classes themselves represent parts of other classes. They group together properties or relationships used in the definition of their "parent". The composition relation is used to represent these kinds of parent-child.
But among the classes in the model there are some that are not used like this. These classes represent concepts that can stand on their own, are not use to describe part of a larger concept. These we will call "root entity classes". In the model they can be identified by the fact that neither they, nor any of their sub or base classes are part of another class, a child in a parent-child relation.
These are the classes that represent the model's core concepts and their identification is a first important choice in the modelling effort. In the current model there are actually two separate collections of classes that are root entities. The Party class represents an individual or organisation. It is used for indicating who/what wrote simulation codes or ran simulations. The main focus in this document is on the root entity classes in the Resource hierarchy, illustrated in Figure 3.

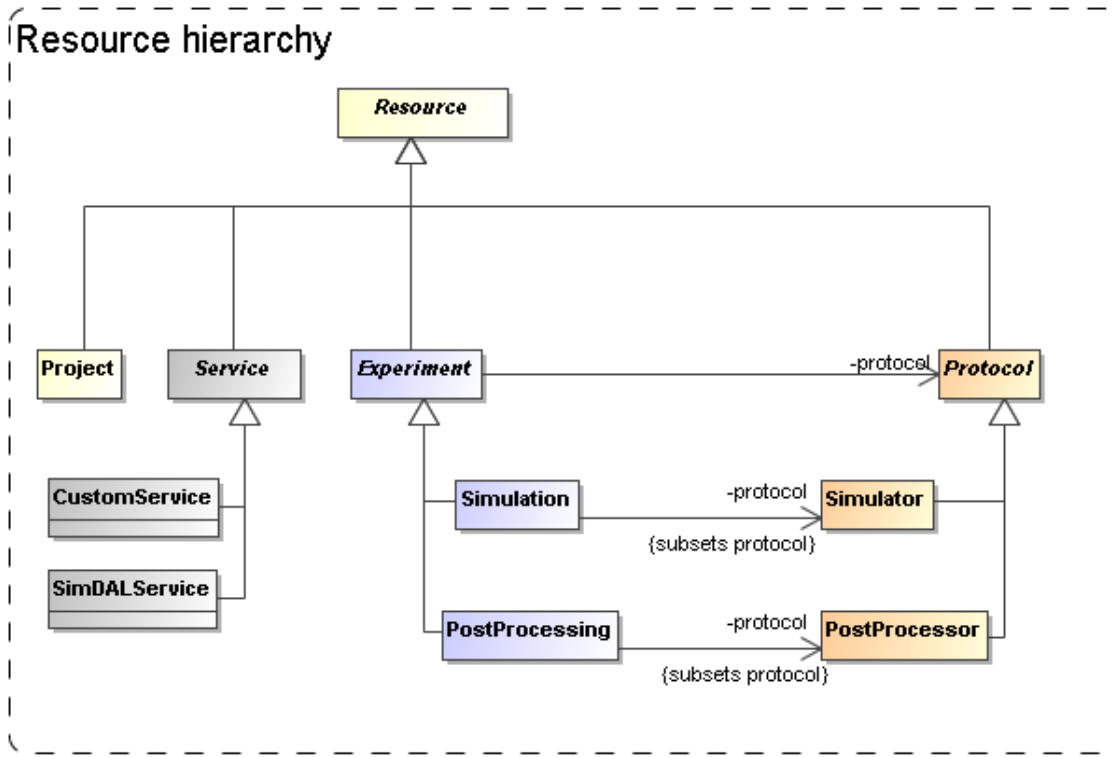

**Figure 3: Root entity classes for SimDM.**

From the top down we start with the ultimate root entity class, *Resource,* which defines components common to all the main classes. The layer below it contains *Protocol*[5]*, Experiment*, *Service* and Project. Protocol is the base class of the concrete classes Simulator and PostProcessor. *Experiment* is the base class of Simulation and PostProcessing. *Service* is the base class of CustomService and SimDALService. Project has no subclasses and is concrete.

Our choice for the root entities follows the domain model in concentrating on the scientific experiments as a whole. The *Experiment* class contains, amongst other components, classes representing the actual results (represented by the OutputDataset class) that people may wish to access. Those are *not* the core concepts in our model. This is in contrast for example to the spectrum data model [11] which focuses on the representation of the spectrum, and has the provenance and other metadata as sub-components.

One reason is that an experiment can exist without having (yet) produced any results, but to have results (as defined here) one always needs an experiment. This is a clear example of a parent-child dependency, where the child's life-cycle depends on that of the parent. The standard way to model such relationships is using a composition relation and that is how we have modelled it. More about the way we model results below in 3.7.

---

[5] The use of the name Protocol for the concept we introduce here has led to comments by some reviewers who feared confusion with the use of the same name in for example DAL protocols. In this document, the capitalised term Protocol will refer to the class in the model. When confusion with other usages might arise we may use the phrase "(experimental) Protocol".

The separation between *Protocol* and *Experiment* is an important feature that we directly take over from the domain model. This design was already motivated in Section 2.2 and is related to the Measurement-Protocol pattern in [27]. That pattern says that when one does a *measurement* (of some property) it is important to remember the *protocol* by which the measurement was made ([27], p65). In [12] this was extended to experiments, which in general consist of large numbers of "measurements", all done in similar ways. Whereas the term measurement seems to be more applicable to observations, it is simple to generalise the concept a bit and apply it to the *calculation* of properties during a simulation. Actually this is similar to the CalculatedMeasurement in [27],

An important reason to keep this separation between *Experiment* and *Protocol* also in our logical model is to avoid having to redefine the parameters and other aspects of a simulation code each time a simulation is run.

The *Service* class did not appear in the original domain model in [12], but we introduced it in the model in 2.2 under the name WebService. In our model it represents a way to provide access to results of experiments. We could have tried modelling the way results are stored in files etc., but deemed it too complex to do so. This is in contrast for example to the spectrum data model, where we can model the data directly and even can predefine the representation of the data. There an access reference to the data files can be given to download a result. For simulations this is in general not possible. In many cases simulation codes have their particular proprietary formats, often storing single results over multiple files. Hence we merely allow users to describe services by which one can access results. Here we only make a separation between custom services and services following any DAL service specification that is under construction in the IVOA.

The Project class represents a scientific project, acknowledging that these in general use one or more experimental protocols to perform multiple experiments. This class is introduced to allow for example publishers to group simulations and post-processing runs that were produced with a common goal. It was inspired by a discussion on whether some of the SimDM/Resources could be registered as Registry Resources as well[6]. Many of the simulations registered in a SimDB will not qualify for the same reasons that individual images do not qualify to be registered. Resources in an IVOA compatible registry are relatively coarse grained; correspond to archives full of images published through a SIAP service for example. A Project can be used to define such collections also in SimDM. And indeed one may wish to register such collections separately in a registry.

In a data model one can use aggregations of the corresponding concepts to build such relations. In fact we have the single aggregation between Project and *Resource*, providing the user the freedom to include *Service*-s and even other Project-s.

The root of the hierarchy of entities is formed by the *Resource* class. This class is introduced as a convenience to hold on to information common to all its sub classes. Its name is obviously inspired by the Registry's Resource [13] and it also holds on to curation information. It "is not a" Registry Resource though in the strict OO modelling sense. For example it does not inherit all features of that class. But this is mainly because, as mentioned above, most SimDM Resources will not qualify as Registry Resources.

---

[6] Thanks to Ray Plante for his contributions to this discussion.

## 3.3 Object types: real and simulated

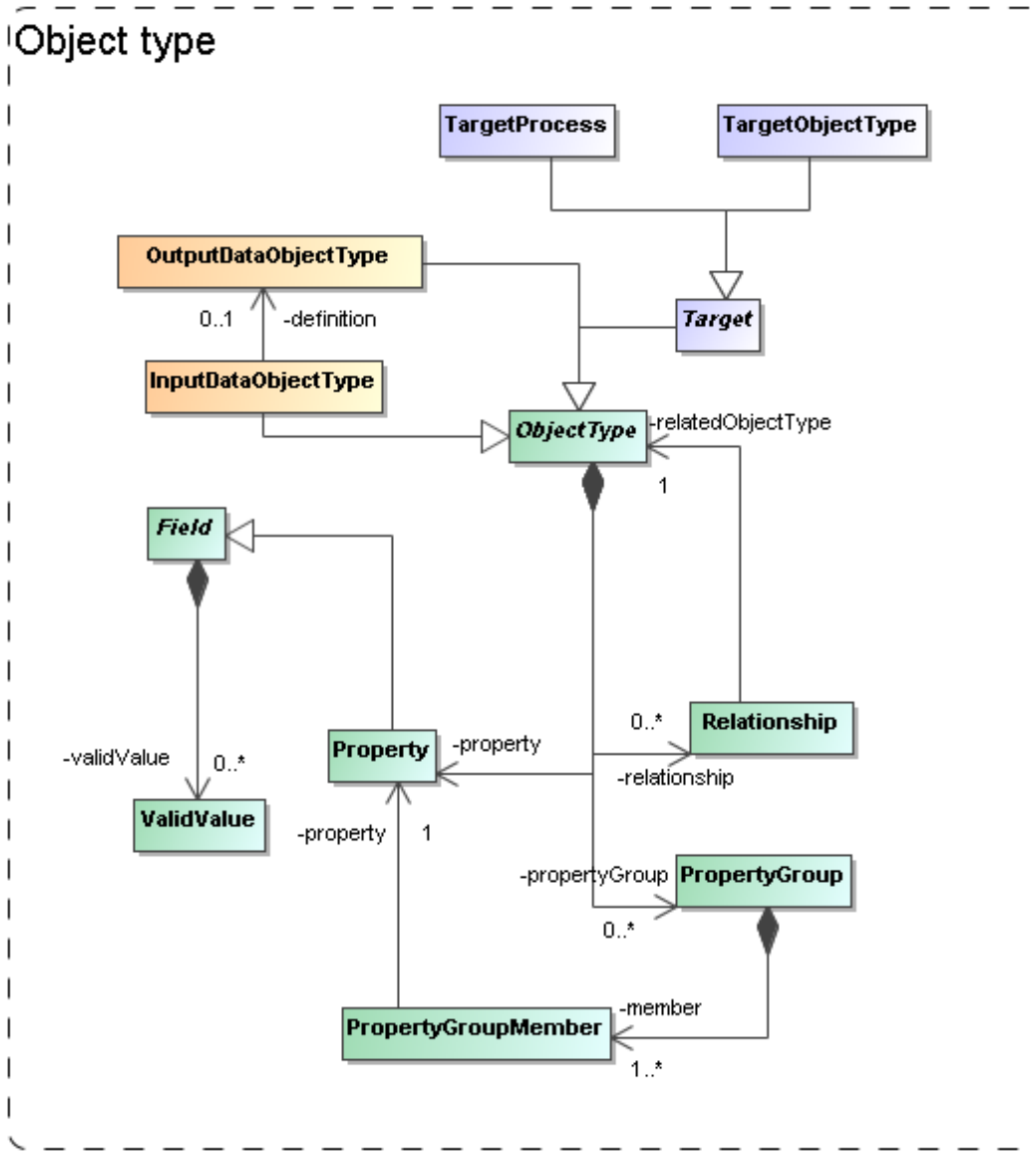

**Figure 4: The model needs to describe the types of objects that are being simulated/used. We model this in quite some detail in a hierarchy of object types, with properties, grouping of properties and child objects corresponding to nested objects.**

In various places in the model we need components that allow the user to describe complex (data) objects explicitly. For example, different simulation codes represent different parts of the world in different ways. And hence they produce (and use) different types of data. For example Adaptive Mesh Refinement codes represent a part of the universe using a grid of cells of varying sizes, whereas a Smooth Particle Hydrodynamic code uses particles with extent. Hence the *Protocol* class needs components for describing the building blocks of the model world it represents (OutputDataObjectType in Figure 4).

And the same code can often be used to represent very different types of astronomical objects. E.g. a particular AMR code might be used to model the internals of a galaxy cluster as well as a supernova explosion. To be able to describe such objects with some more detail [7] (say mass of the cluster) *Experiment* needs a way to describe physical objects that are produced by a run, (Target ObjectType and -Process in Section 3.6).

We abstract the building blocks required to describe such a model world as *objects* with *properties* and *relations*. In SimDM we support this with a limited version of an object oriented meta-model, directly inspired by the UML profile in Appendix B. We add a package object that contains the model elements that will be used in other parts of the model.

The core concept is the abstract *ObjectType* class. An *ObjectType* contains a collection of Property-s that corresponds to the simple attributes used to describe an object. Property is a subclass of *Field* which defines its main attributes such as name, description and data type. Also a *Protocol's* InputParameter (see 3.5) is a *Field*, similar to the way a VOTable's PARAM and FIELD share a common structure. In Figure 5 we show an example instance diagram illustrating the use of OutputDataObjectType (a subclass of ObjectType) and their properties. A similar diagram could in principle be built for images with pixels, or AMR codes with grid cells, etc.

Another similarity with the VOTable structure is the possibility to group Property-s in a PropertyGroup. An example where the latter may be useful is in grouping all properties related to the position of a particle, or the chemical abundances of an AMR grid cell[8].

---

[7] One of the most important questions to be asked of simulation database, see the list in 2.1.
[8] One could argue that these examples could (should) themselves be represented as types, similar to the value types in the UML Profile (see Appendix B.5).For example a Position type might be more suitable than a grouping of (x,y,z) properties. We have refrained from adding this feature in the current version of the model.

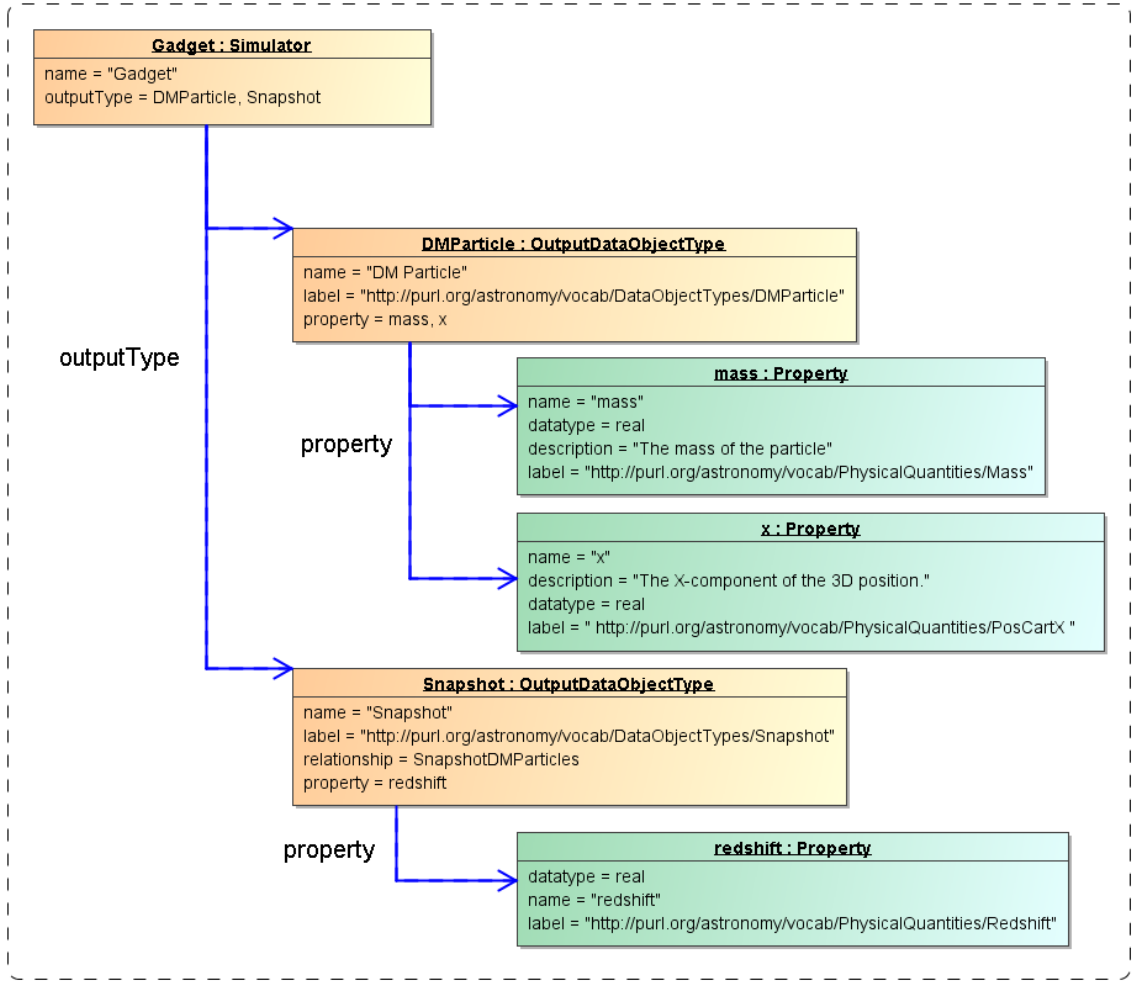

**Figure 5 Example instance diagram showing a Simulator with two output types, "Snapshot" and "DMParticle", each with properties. Note the SKOS concepts used in the label attributes. These may not (yet) exist in the corresponding vocabularies.**

To describe (hierarchical) relations between different objects an *ObjectType* has a collection of Relationship-s, which can be used to define aggregation, composition or reference relationships between different *ObjectType*-s. The instance diagram in Figure 6 shows the same Simulator and output types as in Figure 5, but adds the relationship of type "composition" between the types "Snapshot" and "DMParticle". This relationship represents the definition of a snapshot as a collection of particles.

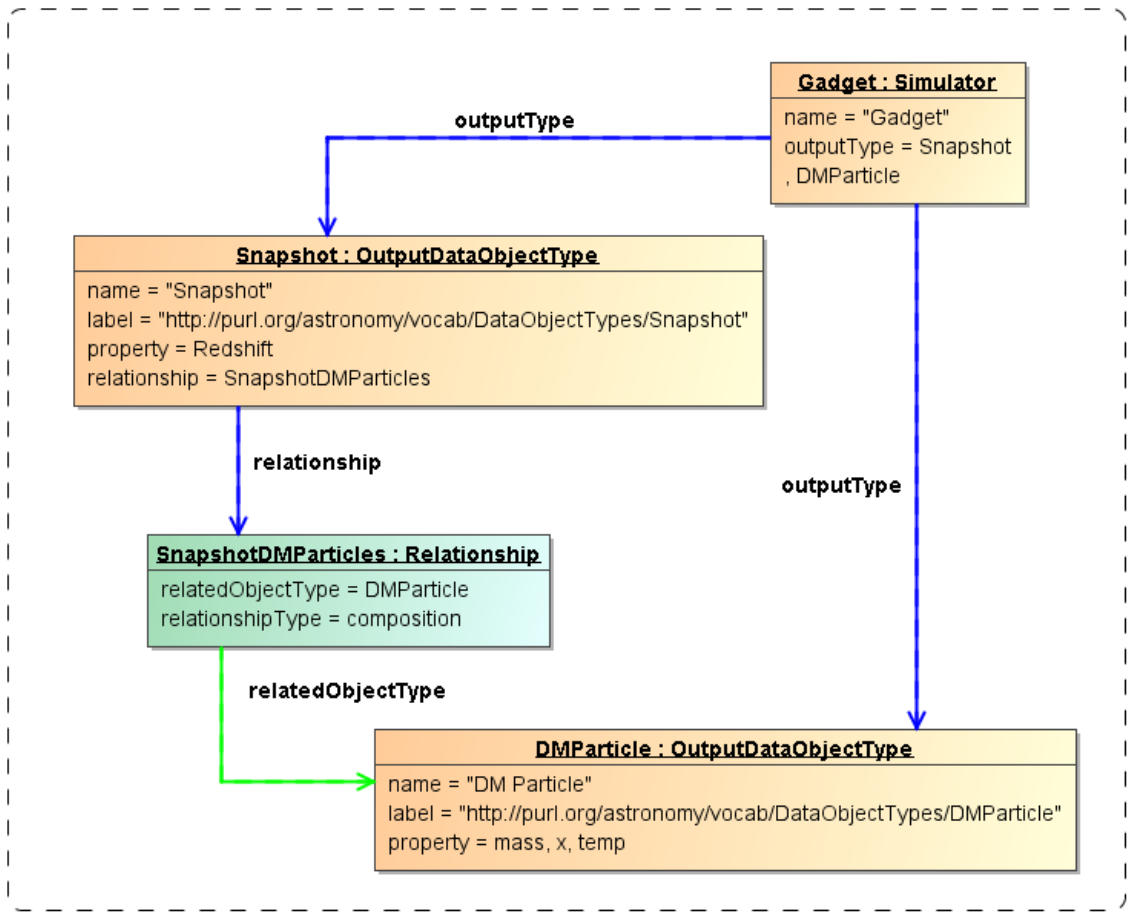

**Figure 6 Example instance diagram for an object type hierarchy defined on a protocol.**

## 3.4 Physics, Models and Algorithms

An important characteristic of simulation codes is what physical systems and processes can be modelled and how these are represented in the program. The Simulator class represents computer codes that create numerical models of the world. Simulators do so by representing physical processes using numerical algorithms that act on model representations of real world objects. In our model, see Figure 7, physical processes are represented by the Physics class. It is contained in the Simulator class, not in the more general *Protocol*. In effect a simulation protocol is distinguished from other experimental protocols in that it models and implements physical processes.

Physical processes are implemented using particular Algorithms. Algorithms are contained in *Protocol*, as also PostProcessors use them. In that case they implement the processing of existing results, and do not model physical processes. Examples of these are particular algorithms for extracting clusters from results of N-body simulations.

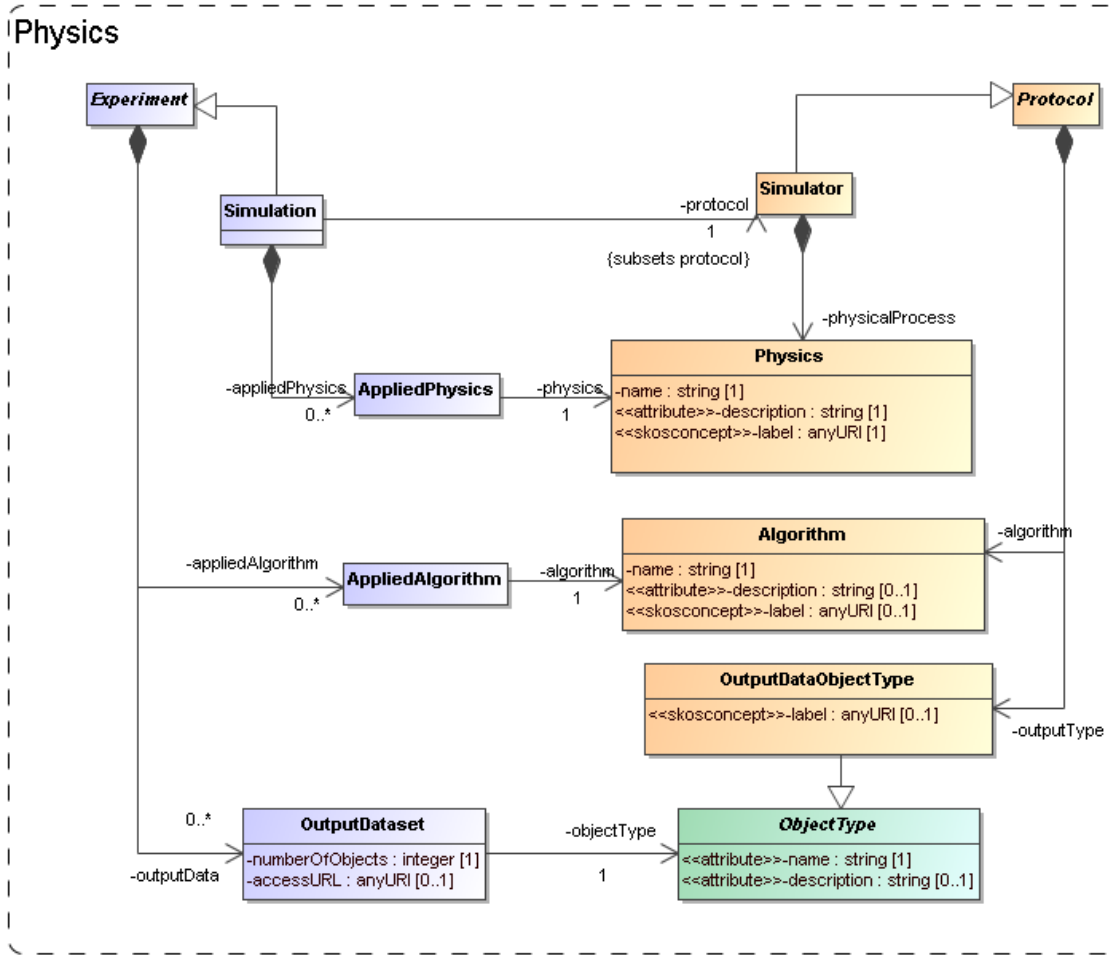

**Figure 7: Modelling the representation of physical processes and objects in Protocols and Experiments.**

Finally, experimental protocols need objects to represent the structure of the physical systems they model. For example, N-body simulations need particles that represent mass moving around. The model uses the OutputDataObjectType for this. This class allows one to define a hierarchy of data objects, from container objects like catalogues, data cubes or images down to the smallest objects such as particles or pixels.

A complicating factor is that many simulation and other theoretical protocols allow users to choose which physical algorithms to use, or which object types. For example SPH codes such as Gadget can be run in full hydrodynamic mode, or in pure gravity, "dark matter" only. Hence, when defining an experiment, users should be able to indicate the choices they made.

For this purpose Experiment and Simulation have collections of classes that allow explicit links to the components they use off the Protocol. The collection of AppliedPhysics references the Physics used, AppliedAlgorithm the Algorithm and OutputDataSet references the ObjectTypes used. In contrast to the former two, OutputDataSet has more structure that will be discussed below.

## 3.5 Parameters: definition and values

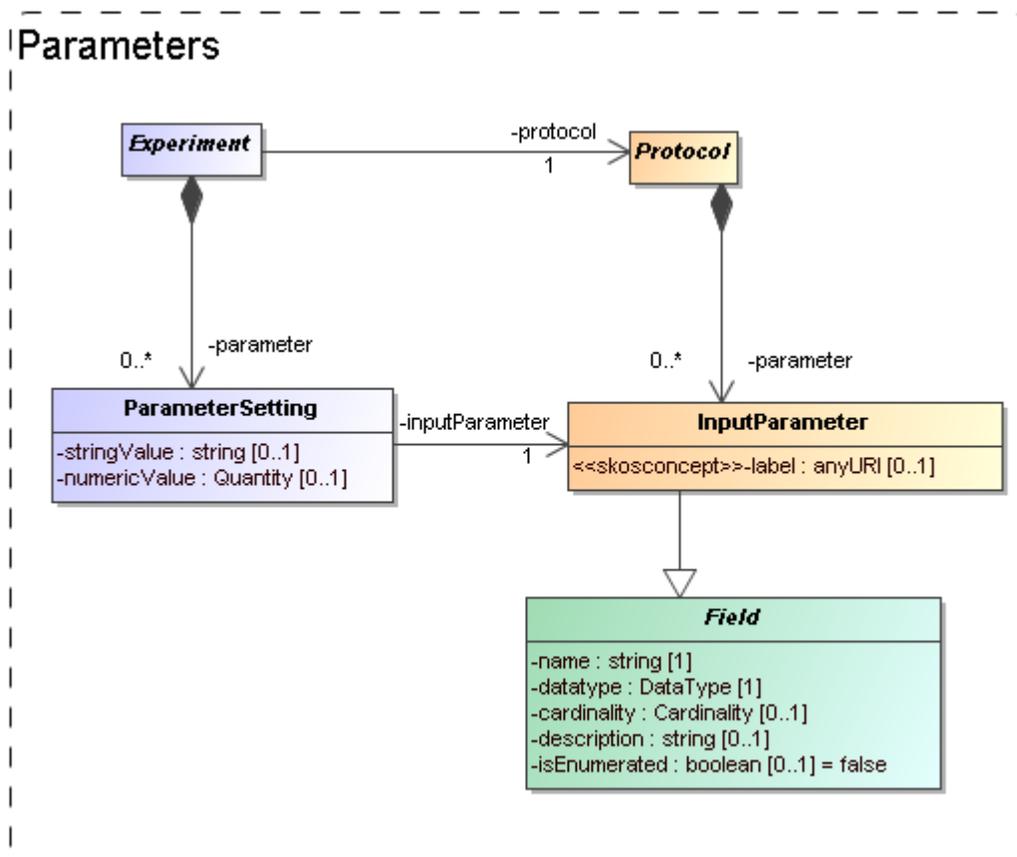

**Figure 8: Modelling the parameters: definition under (*experimental*) Protocol, values under Experiment.**

Software codes generally require some level of configuration before they are executed. In many cases this translates into a collection of parameters that must be given values. The parameters are defined by the code and we model this by an InputParameter class that is contained by *Protocol*. Assigning values to these parameters however is the responsibility of the experimenter and is explicitly modelled as a ParameterSetting class contained by *Experiment*.

Input parameters are defined by the attributes name, datatype, label and other properties familiar for example from the PARAM field in VOTable[9]. Most of these are inherited from the *Field* class, which will be discussed in Section 4.1 below.

Because the details of the parameter are defined on the InputParameter class, the ParameterSetting needs only a pointer (the inputParameter reference) to the appropriate input parameter and a value. A problem for this model though is what data type to assign to a possible value attribute. We have no knowledge in advance on the data type of the input parameter for which a value is set. This is only known at the instance level, not at the model level. We do not know whether a certain parameter will be integer, or real, or maybe a string. Our current solution is to allow two different representations of a value, namely a numericValue, of type real and a stringValue of type string.

---

[9] We generalize the ucd attribute on VOTable's PARAM and FIELD to a label attribute with stereotype <<skosconcept>>.

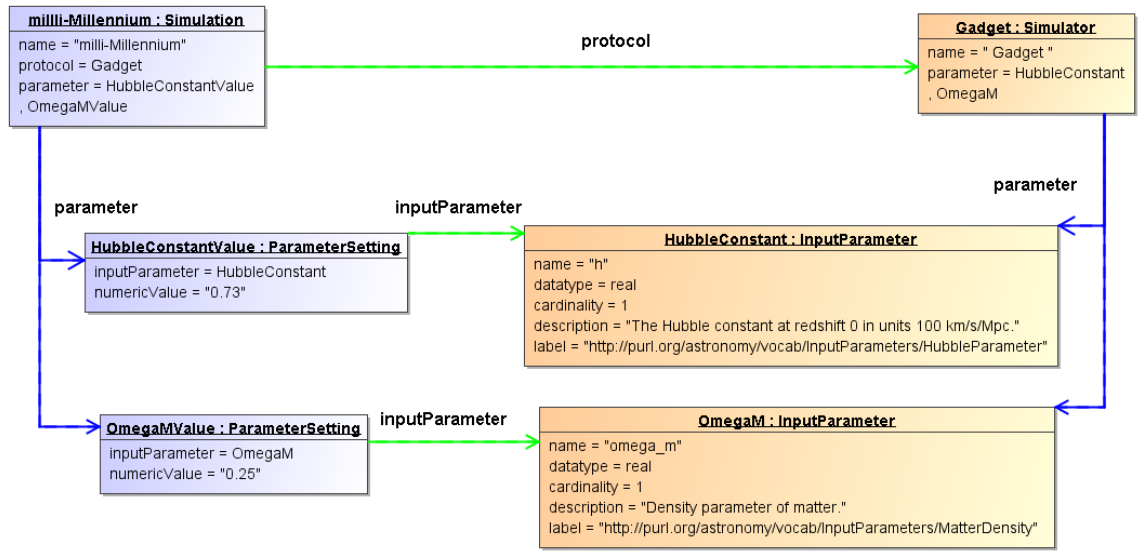

**Figure 9 Example instance diagram of parameter definitions and value assignments.**

Figure 9 shows an example instance diagram for parameter definitions and value assignments. It shows the Simulator "Gadget" that defines two input parameters: "h", representing the Hubble parameter and "omega_m", representing the density parameter for matter ($\Omega_m$). Note that these parameters have been labelled using the appropriate SKOS vocabulary. The figure furthermore shows a Simulation run with Gadget (as indicated by the protocol reference between the objects) which assigned the values 0.73 and 0.25 to "h" and "omega_m" respectively. Here the inputParameter references indicate which value was assigned to which parameter. One also sees the advantage of the normalized design, with the parameter definition separated from the parameter value assignment: The bulky definition of an input parameter, with a potentially long definition, need not be repeated; instead it is reused by every Simulation that uses the same Simulator.

## 3.6 Target: Goal of experiment

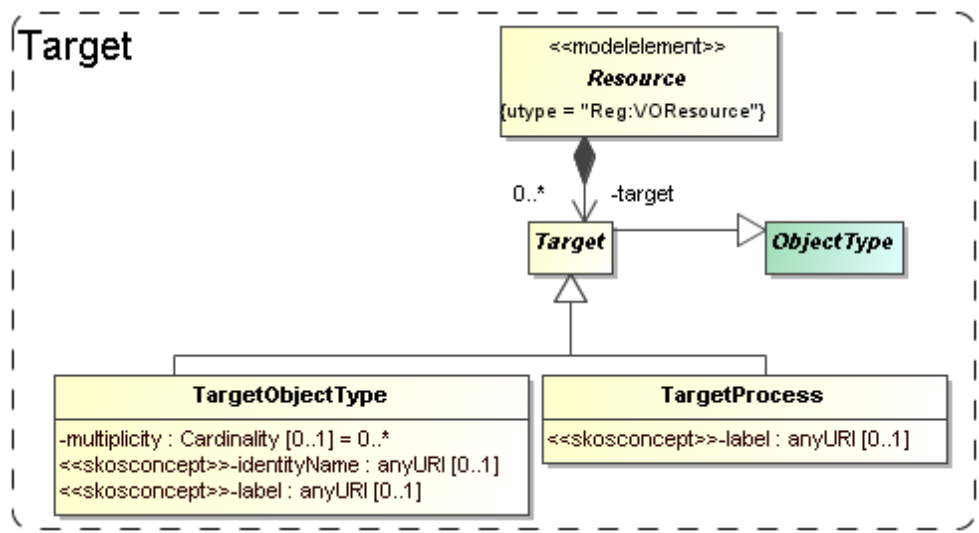

**Figure 10: Modelling the goal, or target of a generic resource as objects and/or processes.**

Generally the first piece of information that the scientists we polled were interested in regarding simulations was *what* was simulated. I.e. what type of object: a galaxy merger, a galaxy cluster, the large scale structure of the universe? This information in general says something about the goal that the scientists running the simulation had.

In certain cases the simulation code itself may completely prescribe the type of objects and physical processes that are modelled. As example take population synthesis models such as the Galaxev library[10], producing spectra of galaxies.

But many simulation codes allow many different types of objects to be modelled, and even allow one to vary which processes are actually modelled. Also in many cases the actual object that is being simulated is not an intrinsic property of the simulation code, but is a derived property of the actual simulation. For example an N-Body code in general does not contain "galaxy particles". But one can use it to follow the evolution of millions of low mass particles that are in a particular configuration that together model a galaxy. But it can also be a globular cluster, or a filament in the large scale structure.

To cover the concept of the target of an experiment or protocol, or the goal of a project, we add two classes, TargetObjectType and TargetProcess. A TargetObjectType represents an object, or a physical system in the real world, such as a galaxy, a star etc. TargetProcess represents a physical process such as gravitational clustering or turbulence. This recognises the fact that some simulations are run with the goal of investigating a process, rather than producing a model of a physical system.

Both these classes are subclasses of *Target*, which itself is again a subclass of *ObjectType* defined in the next section. *Target* is contained by *Resource* so that by inheritance they are available to all sub classes. We do not model the *Target* objects in full possible detail. That we leave to future astronomical ontologies. We restrict ourselves to a description (inherited from *ObjectType*) and a semantic label attribute which identifies the intended concept using a standardised name from a SKOS vocabulary (see 5.2 below).

---

[10] Bruzual and Charlot, 2003: http://www.cida.ve/~bruzual/bc2003

## 3.7 Results: data sets and their statistical summary

We assume users of a Simulation Database will want to gain access to results of simulations and related experiments. This is the same as we assume of users of Simple Image Access or Simple Spectral Access services. For those services the user knows what to expect, a FITS image in one, a spectrum serialised according to the spectrum data model in the other. I.e. for those IVOA protocols the object types can be (and are) predefined as part of the protocol.

This is not possible for simulations, where we cannot assume to have *a priori* knowledge about the contents of their results. Arguably somewhat simplistically one may claim that images and spectra contain pixels with known properties (space, wavelength, flux). Results of simulations, even when constrained to 3+1D simulations, can contain as their *fundamental constituents*: point particles, particles with size and structure, mesh cells of fixed or varying size, Voronoi cells[11], structured halos, galaxies, radiation fields, galaxy merger trees etc. And any of these object types can come with any collection of properties: position, velocity, mass, temperature, chemical composition, entropy etc.

Precisely for this reason users will want to gain knowledge about the contents of simulation results to decide which simulations might be of interest to them. Hence the model must support description of the results explicitly.

Figure 1 illustrates how this is achieved in the domain model: Experiments produce Results that consist of ObjectCollections of Objects (pixel, N-Body particle etc) of a particular ObjectType. The ObjectType defines the structure of Object as a collection of Properties (position, velocity, flux etc), and an Object, being an instance of the ObjectType, assigns values to these properties. Which ObjectTypes and Properties are available is defined by the *(experimental)* Protocol according to which the Experiment is run.

SimDM deviates a bit from this model, for example the separate Result class is absent and replaced by a hierarchy of object collections. SimDM also adds a feature to this model that is directly inspired by the Characterisation Data Model [16]. In many cases a listing of the actual data objects, using the Object and ValueAssignment classes would be overkill. For example consider the case of N-Body simulations. We might define two object types, one representing the particles, another representing a snapshot, corresponding to the state of the simulation at a given time. Whereas it might be useful to contain individual snapshot objects in a meta-data repository of simulations, clearly it is not useful to list all the particles that make up these snapshots. It might be useful though to provide some statistics of these collections and just such a feature we have added to the model.

---

[11] http://www.mpa-garching.mpg.de/~volker/arepo/

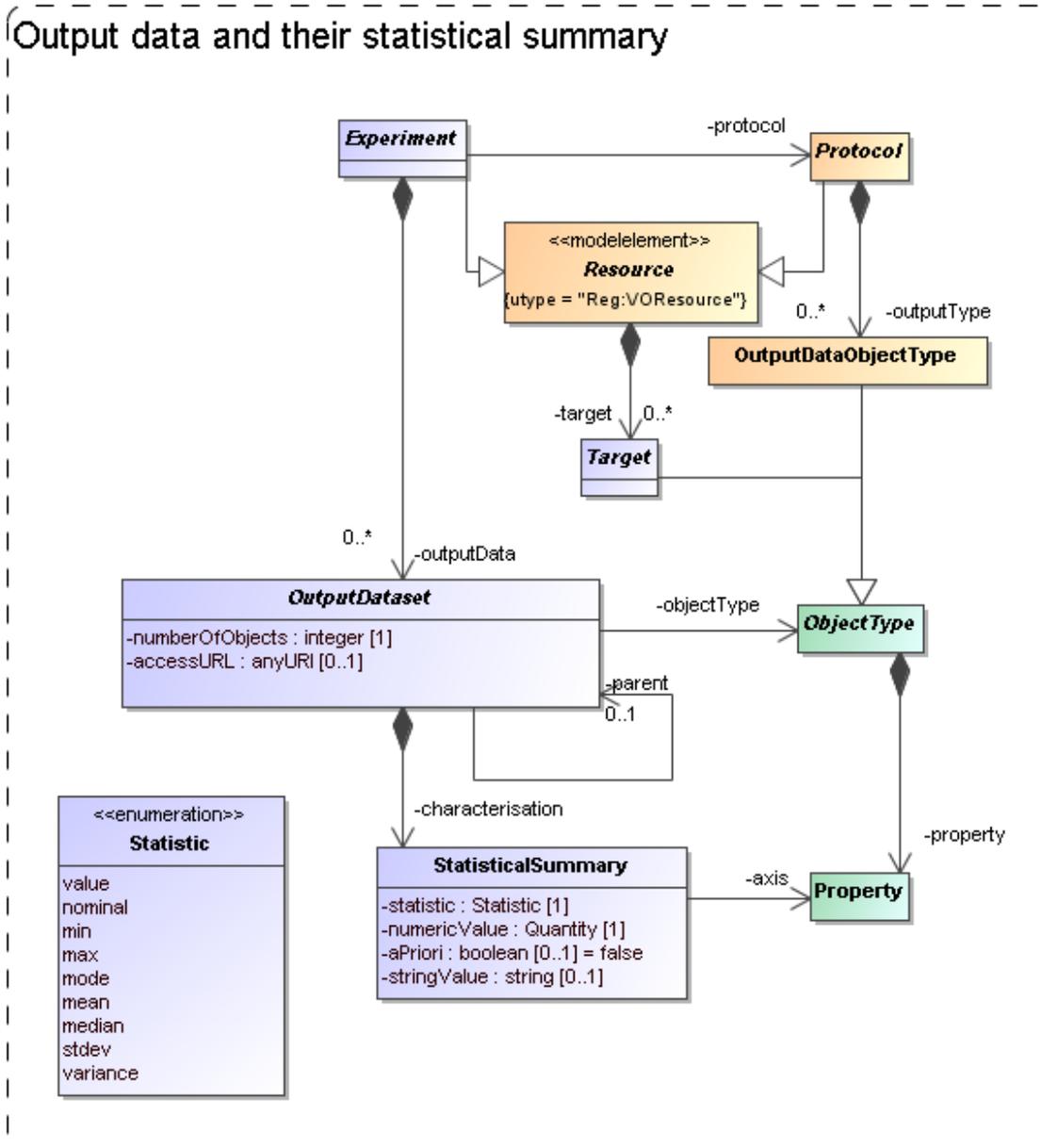

**Figure 11: Modelling Output data sets and their contents.**

The model for results and their contents in SimDM is shown in Figure 11. Results of experiments are represented by OutputDataset-s. As described in Section 3.2 above, OutputDataset represents a *collection of objects* of a particular *ObjectType (*indicated by the objectType reference) produced by the *Protocol* during the *Experiment*. In the typical case the referenced object will be an OutputDataObjectType defined for the given *Protocol*. But by referencing the base class ObjectType, an OutputDataSet may also represent a collection of TargetObjectType-s (see 3.6). Hence a user has the possibility to characterise the astronomical objects that have been simulated as well as the more technical objects produced directly.

The OutputDataSet class indicates which *types* of objects are used in a particular Experiment. The class contains two collections to provide more detailed information on the actual results. The first is most straightforward: Users can add DataObject-s to an OutputDataSet. These represent direct instances (hence data *objects*) of the ObjectType referenced by the OutputDataSet. These objects contain PropertyValue-s that assign values to the Property-s defined on the ObjectType.

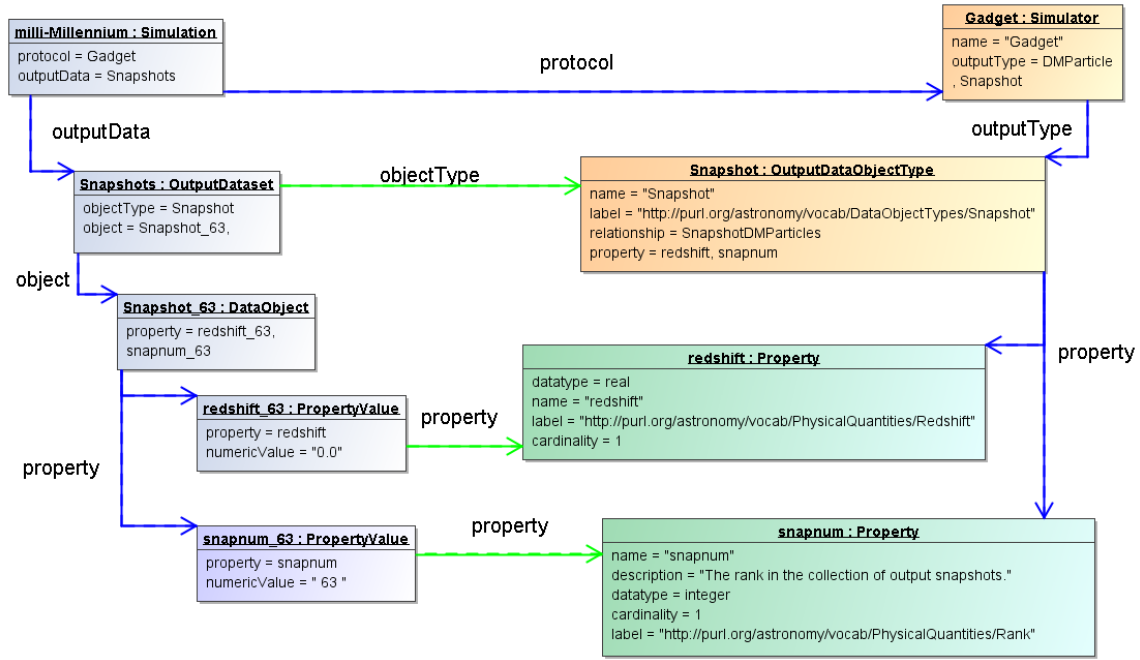

**Figure 12 Example instance diagram illustrating the use of the DataObject and PropertyValue classes.**

This part of the model is illustrated in Figure 12 using an instance diagram. The "milli-Millennium" simulation has an OutputDataset of objectType "Snapshot". This data set contains one DataObject, with a PropertyValue for each of the two Property-s defined on the OutputDataObjectType: "redshift" and "snapnum".
The DataObject has collections of ObjectReference and ObjectCollection that can be used to represent explicit relationships between objects or between objects and object collections respectively. They implement a Relationship defined on the DataObject's ObjectType. An ObjectCollection object links the DataObject to another OutputDataSet, representing a composition relation. The precise Relationship that is implemented is indicated by the collectionDefinition reference. Similarly an ObjectReference indicates a reference relation between one DataObject and another.

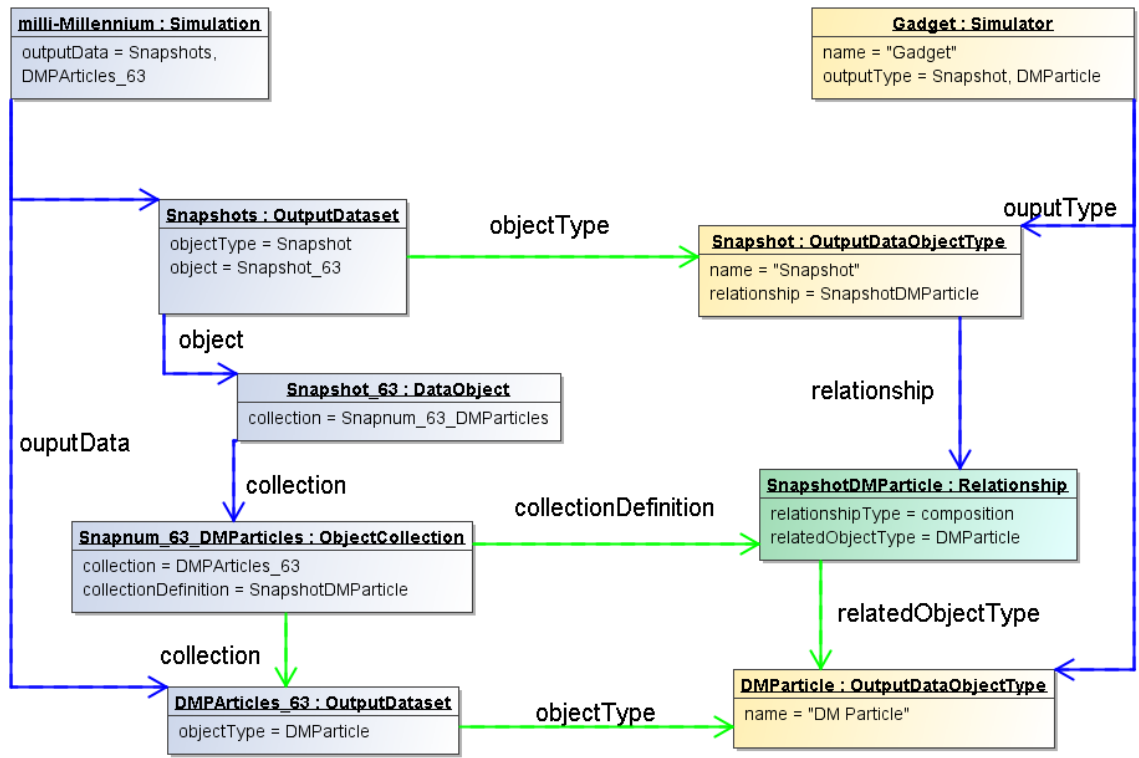

**Figure 13 Example instance diagram illustrating the implementation of object relationships on an experiment's result objects.**

This somewhat complex part of the model is illustrated in Figure 13.

The second way to provide more detailed and quantitative information about object collections is statistical. For many cases a complete listing of all data objects in an OutputDataSet is not feasible, especially not in a repository aimed at storing metadata about simulations such as SimDB. But it may be useful to provide summarising information about the objects in the data sets.
To support this use case, the model contains the StatisticalSummary class which allows users to assign statistical values such as a mean or a min/max value to Properties of the OutputDataset.objectType. Which statistic is used is described by the statistic attribute. An illustration of the statistical summary is given in the instance diagram in Figure 14

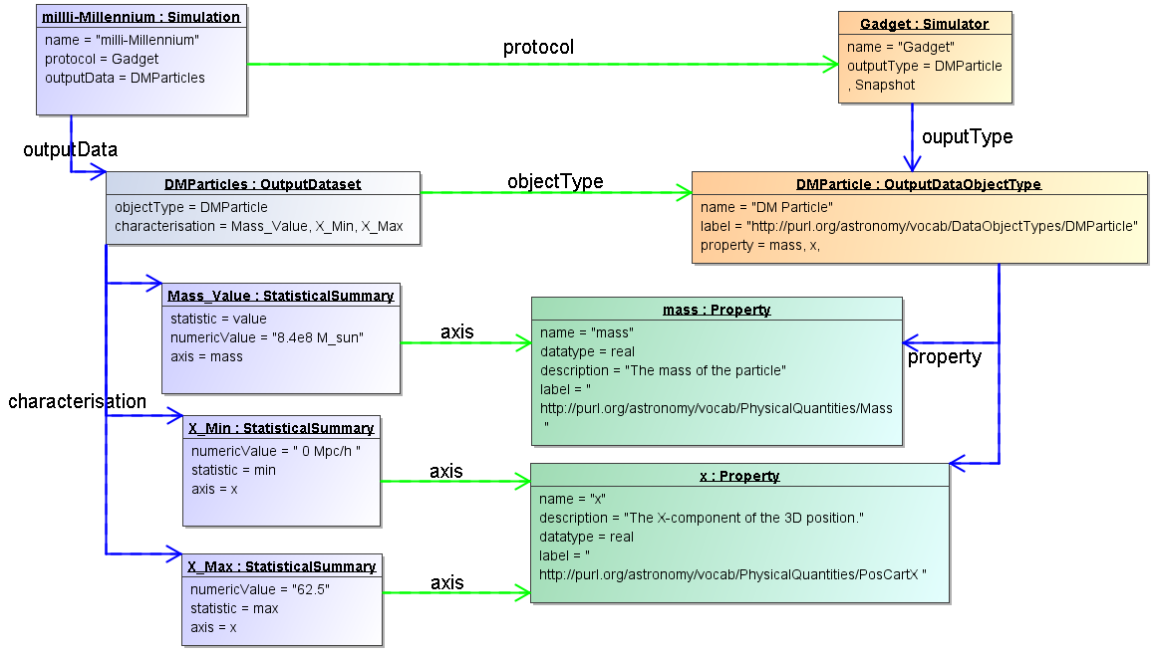

**Figure 14 Example instance diagram of characterising a result using the StatisticalSummary class.**

The example shows an OutputDataSet of objectType "DMParticle". That object type defines properties "mass" and "x". The simulation has a dataset of particles and declares that all particles have the same mass 8.4e8 $M_{sun}$/h. It does so by assigning the statistic "value" to the StatisticalSummary object referring to the "mass" Property. Similarly the min/max values of the "x" property are declared to be 0/62.5.

Extensions of this statistical summary to more detailed summaries such as histograms can be easily imagined, but have been left out of the model as they will have less relevance for discovery, which is the main use case for the model.

One further feature is important and is represented by the boolean **aPriori** attribute. This attribute describes whether the statistic that is used in the summary is an *a priori* or an *a posteriori* statistic. An *a posteriori* statistic is calculated using the results after they have been obtained during the running of the experiment. For example an *a posteriori* mean will likely correspond to the usual expression,

$$\frac{1}{N}\sum_i^N a_i \, ,$$

where the $a_i$ are the values of some property.

In contrast *a priori* statistics characterise the possible values of the observables *before* the experiment is run. In certain cases *a priori* knowledge is available that restricts the possible values that certain properties may obtain in an experiment. An example is a lower bound set on the number of particles that a cluster must contain to be included in the result of a cluster extraction of an N-Body simulation. This can be indicated by a StatisticalSummary object with statistic=min and aPriori=true.

Knowledge about the *a priori* statistics is important in the interpretation of the results. In the previous example, when interpreting the mass multiplicity function of a cluster catalogue extracted from an N-Body simulation, it is clearly important to know what the lower limit was on the mass of clusters.

In general *a priori* statistics are the result of, and may often be derived from the input parameters. However this derivation may not be obvious and will in general require intimate knowledge of the parameters of a *(experimental)* protocol. The *a priori* statistic may then facilitate the discovery of catalogues that should contain halos of a certain mass.

## 3.8 Data access services

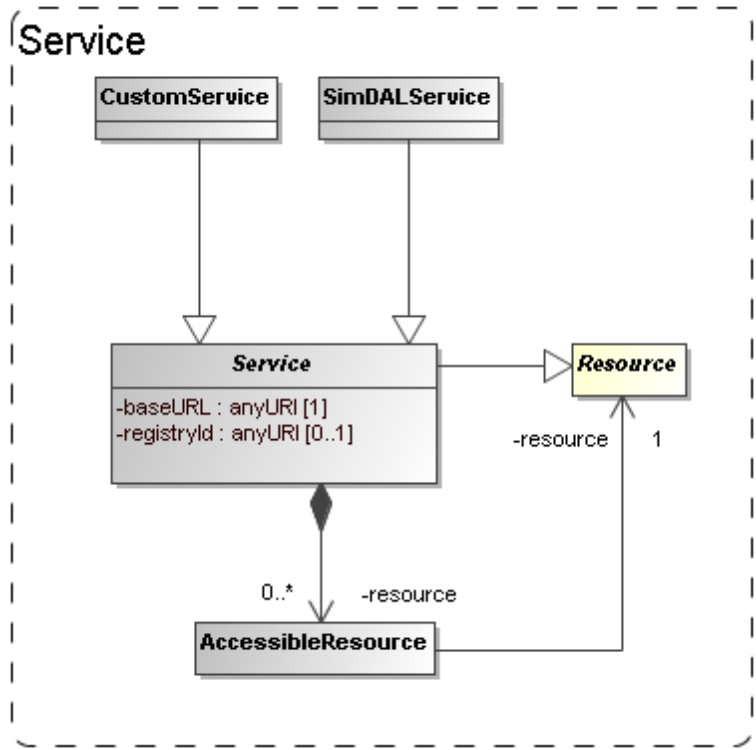

**Figure 15: The model for (web) services giving access to resources.**

The goal of the Simulation Database is to allow scientists to find simulations of possible interest. Once these are found the question is what can be done with them. Clearly knowledge of their existence will be useless if the researcher will not be able to somehow gain access to the results. The usual way this is done in the IVOA, for example in the simple image and spectral access protocols, is that the result of a discovery query contains an access URL that may be used to download the actual image or spectrum, where moreover the format of the returned resource, FITS, VOTable or XML document will be known beforehand.

It was perceived from the beginning even that for the type of simulations that were supposed to be described a simple download would be unfeasible simply based on the size of many of the typical N-Body or AMR simulations. This assumption still holds and any DAL protocol for simulations must be designed to define special purpose services for retrieving parts of such simulations for example.

Also in the data model we want to indicate how the relation is between the results and services. This part may be used in the SimDB specification to allow users to register services and the Resources they give access to.

In the model the *Service* class, already introduced in 3.2, represents such access services (see Figure 15). This class can be explicitly linked to the Resources it gives access to through a collection of AccessibleResource-s. The class has concrete subclass SimDALService. This represents services providing access to the results of a limited set of *Protocols* through a DAL protocol. The CustomService is introduced so that users can also publish non-standard services. This part of the model is still rather summarily treated and may need to be updated depending on developments in the DAL specification.

# 4 Serialisations

According to policies of the data modelling working group, first decided in Cambridge, 2003, a data model should be presented using a UML diagram, a corresponding XML schema and a list of UTYPEs. We have created these both using rules that derive the products directly from the XMI serialisation of the UML data model.

## 4.1 SimDM/UTYPE

The original goal of the data model presented here was to define the structure of a relational database supporting the SimDB specification. A first draft of a note proposing that spec can be found in [22]. SimDB will use TAP [10] to define the IVOA protocol for querying this database using ADQL [24]. The results of such queries will be tabular and serialised as VOTables. Such a VOTable will contain a filtered subset of the information in the database, but in general in a different form compared to the structure of the data model. To indicate the meaning of data elements in such a VOTable, the IVOA has invented the concept of UTYPEs.

A UTYPE is a "pointer into a data model"[12]. The VOTable XML schema implements this concept as attributes on various elements, e.g. FIELD and TABLE and many other elements. The value of such a UTYPE attribute should identify an element in a data model that is represented by the element itself. For example a table might point to a class definition in a data model, and a column (FIELD) to an attribute.

It has become common practice to provide for an IVOA data model a list of UTYPEs. The Spectrum data model (see [11]) was the first to add explicit

---

[12] See 5.3.3 for our position on the discussion that is still going on regarding UTYPEs.

UTYPE-s for each of the attributes in its model and the Characterisation data model [16] has followed that example. We follow these examples by assigning UTYPE-s explicitly to all elements in the model.

Our goal was not to have to make this a separate effort, but if possible to generate the list of UTYPEs directly from the model. Our goal was to assign UTYPEs to all identifiable elements in our model and these should be unique.

To this end we define a set of production rules phrased using the special names in our UML profile. We have made a guess as to what the format for UTYPEs will be. In the previous data models a UTYPE was made of a word consisting of dot-separated "atoms", similar to UCDs, but without the ";". We use a slightly different format to make the distinction between different syntactic elements from the profile somewhat clearer and also to guarantee uniqueness of each UTYPE within the data model context. Once (if?) a format is settled on within the IVOA we will easily be able to adjust our definitions.

The important point we want to make is that it is possible to define simple rules that can automatically produce *unique* UTYPE-like words for all elements of a data model, i.e. the only discussion that may be required is on the rules for doing so IF a fixed format is preferred (see Norman Gray's ideas[13] on why this might not be necessary).

The following BNF-like expressions define the particular rules we have used for deriving the UTYPEs from the UML model:

```
utype              :=   [model-utype | package-utype | class-utype |
                        attribute-utype | collection-utype |
                        reference-utype | container-utype
model-utype        :=   <model-name>
package-utype      :=   model-utype ":/" package-hierarchy
package-hierarchy  :=   <package-name> ["/" <package-name>]*
class-utype        :=   package-utype "/" <class-name>
attribute-utype    :=   class-utype "." attribute
attribute          :=   [primitive-attr | struct-attr]
primitive-attr     :=   <attribute-name>
struct-attr        :=   <attribute-name> "." attribute
collection-utype   :=   class-utype "." <collection-name>
reference-utype    :=   class-utype "." <reference-name>
container-utype    :=   class-utype "." "CONTAINER"
identifier-utype   :=   class-utype "." "ID"
```

For the SimDM these rules produce a list of UTYPEs for the model. For each model element we provide the UTYPE in the HTML documentation in [5] and we provide a complete list at the end of that document[14]. Note also that a URL of the type

    `<URL-to-HTML-doc>#<utype>`

will link one directly to the documentation for the corresponding data model element. This is in conformance with a suggestion made by Norman Gray[13].

---

[13] http://nxg.me.uk/note/2009/utype-proposals/
[14] http://ivoa.net/Documents/SimDM/20120503/html/SimDM.html#utypes

When representing components of the data model in a VOTable (for example), these UTYPEs SHOULD be used, in particular when the VOTable contains results of ADQL queries to a SimDB/TAP implementation (see *SimDB Services*). Alternative views and representations of the SimDM, for example in any DAL protocol for simulations, SHOULD use these UTYPEs to refer to elements in the model.

## 4.2 XML

A specification for an IVOA data model should (must?[15]) contain an XML schema that defines how to serialise data model instances as XML documents. Similar to the case of UTYPEs we did not want to make the design of these schemas a separate effort; instead we want to derive the schema from the model. To do so we have defined rules for relating XML Schema constructs to our UML model. These rules are a completion of those described in [37]. It is based also on a view of what such schemas should look like, restricting the possible set of constructs to be used in schemas representing data models. These design rules have earlier been discussed with and accepted by the Registry and VOTable working groups.

We give here only a short description of these rules. First of all we define two different types of schemas. First we define "type schemas", XSD documents containing only type definitions. For each object type(class) and value type we generate a corresponding `complexType` or `simpleType`. Attributes map to elements of a corresponding data type (simple or complex), collections to elements of a type corresponding to the class. References are harder to represent and will be discussed below.

We next generate a "document schema" containing root elements. The elements in the document schema define the valid XML documents one can write and we choose only "root-entity classes" for their type. That is, only classes at the root of collection trees can be represented as a document. Fragments of these are not allowed. For example, only a complete Simulator or Simulation can be represented in a document, not only a single result, or parameter setting.

Note that this is a choice made for the Simulation Database service specification. The document schema depends on the type schemas through XML schema import declarations. This separation allows flexible usage of the type schemas, for example other services might make a different choice from the types to serve as valid root elements.

The root schema for the SimDM/XSD representation can be found [here][16]. The type schemas and a predefined base schema can be found in the same directory and subdirectories of it. We refer to the *SimDB Services* document for more details on the XML schema serialisation and their use in the SimDB service protocol.

---

[15] See "Rules" on http://www.ivoa.net/cgi-bin/twiki/bin/view/IVOA/HowToParticipate This "decision" was made in the Cambridge 2003 interoperability meeting together with the requirement that data models must be specified in UML.

[16] http://ivoa.net/Documents/SimDM/20120503/xsd/SimDM_root.xsd

Only the mapping of references deserves special attention. Our choice of mapping from UML to XSD elements and our definition of root elements imply that many references must be able to link between different XML documents. For example the *(experimental)* protocol reference[17] in an XML document describing an Experiment must be able to identify a *(experimental)* Protocol that is defined in a different XML document. To do this identification we assume we must rely on an agent that can interpret a serialisation of a reference and use it to look up a corresponding document. Therefore we map references to elements of a particular complexType that we define in a base schema[18]. That same schema defines a type to be used for representing identifiers of objects and the reference serialisation must be able to reproduce such an identifier.

Further technical details of this mapping will be described in the appropriate service definition document.

# 5 Dependencies on other IVOA efforts

IVOA documents are assumed to specify dependencies on other IVOA efforts. We have from the beginning realised that the SimDM effort touches upon various other specifications and general efforts of other working groups [22]. Here we discuss these relations as far as they pertain to the Simulation data model.

## 5.1 Registry

The correspondence between the full SimDB specification and the IVOA Registry will be discussed in the *SimDB Service* note [22]. Here we will address the relation between the SimDM and the Registry Data Model as defined in [14].

---

[17] UTYPE: SimDM:/resource/experiment/Experiment.protocol or
http://ivoa.net/Documents/SimDM/20120503/html/SimDM.html#SimDM:/resource/experiment/Experiment.protocol
[18] http://ivoa.net/Documents/SimDM/20120503/xsd/base.xsd

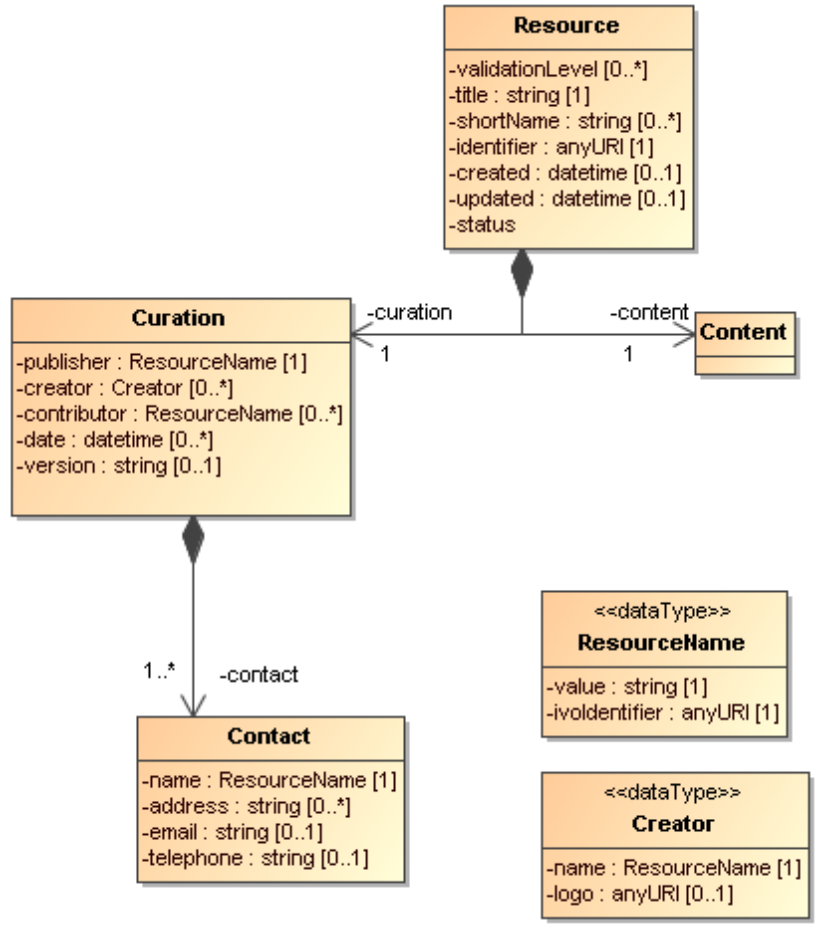

**Figure 16: UML rendering of the Resource complexType from [14].**

In Figure 16 we present a UML rendering of the *Resource* complexType as inferred from the Resource Registry VOResource XML Schema [14]. Comparing that model to SimDM/Resource we can see that these two models for Resource are related, but not identical. In data modelling terms, it is not true that a SimDM/Resource *is a* Registry/Resource (or *vice versa*). *Curation* is modelled differently and arguably with less detail in SimDM, but the main difference is in the *Content*. SimDM provides a very detailed and specialised model for the *Content* of Simulations and related resources, by modelling provenance, motivation and results characterisation. This higher level of detail gives rise to a higher level of granularity in the types of resources stored in a SimDB, which in general will be to fine grained for registration in a Registry. This is similar to the case of a single image, which is not a Registry/Resource, whereas a SIAP-compatible *service*, providing access to many images, is.

A SimDB service itself will have to be registered, i.e. a SimDB service *is a* Registry/Resource. In discussion with Ray Plante (IVOA Interoperability meeting May 2007, Beijing) on this issue it was proposed that some part of the contents could also be registered in a Registry directly, i.e. we should be able to identify Registry/Resource-s in SimDB. Considerations to decide on how to make this identification would be for example that all data products resulting from a well-defined (and published) scientific project could qualify. To represent such a possibility for now we have introduced another subclass of SimDM/Resource: SimDM/Project. This is not much more than an annotated aggregation of other SimDM/Resources, with some additional attributes describing the motivation etc. The metadata of a SimDM/Project is not the same as that of a Registry/Resource, however we propose that we should be able to define a transformation (possibly implemented again in XSLT) to transform a SimDM/Project and produce a Registry/XML representation.

## 5.2 Semantics: Use of SKOS Concepts

In the SimDM, observables, object types, properties, parameters that play a role in a given simulation have to be defined explicitly, for the world of simulations is too large to define all possibilities explicitly in the model itself. This is in contrast for example to the spectrum data model [11] where we know that a flux is determined for a wavelength interval, or a model for images where a flux is determined for a spatial pixel. In principle the publisher of a SimDM/Resource has all freedom to name and describe these entities. For other users to understand the meaning of them, we have where appropriate, added an attribute corresponding to a semantic label. This is similar to the situation in VOTable, where FIELD-s can be given a UCD (or UTYPE) that allows users to understand the meaning of a column in the table.

In SimDM we need to generalise this concept as UCDs are not sufficient for our purpose. For example target object types are not covered by the list of UCDs and the same for other elements in our model. The Semantics WG has specified that such vocabularies should follow the SKOS specification [25]. They have also defined a number of such semantic vocabularies in the SKOS format, for example of astronomical objects. We try to anticipate their results by introducing a special type of attribute in our UML profile that corresponds to a concept in a given ontology.

Technically, in the UML profile we have defined a stereotype <<skosconcept>> that can be assigned to an attribute in the UML model. Attributes with this stereotype must define a value for the tag "broadestSKOSConcept".

The intent of this is as follows (thanks to Norman Gray for providing the original text with this formal definition):

<<skosconcept>> attributes take a skos:Concept as their value. In each case, the value is given as a single skos:Concept: such attributes may take any skos:Concept which is a narrower concept than this single typing concept. To be precise, for a typing concept T, any concept c is a valid value for this property, if either:

   c skos:broaderTransitive T

or if there exists a concept X such that

   c skos:broaderTransitive X. X skos:broadMatch T

This just means that, if c is in the same vocabulary as T, then it's connected by a chain of any number of skos:broader, and if it's in a different vocabulary, then there is some X which is in the same vocabulary as c, with a cross-vocabulary link between X and T.

In several cases -- particularly those vocabularies which have been created for SimDM -- there will be a single top concept which everything is narrower than. In other vocabularies -- such as the AstroObject in the thesaurus version of the ontology of object types -- the natural typing concept is not a top concept, or is not the only top concept. This definition also does indicate that it's legitimate for concept c to come from a different vocabulary from T: the fact that c has been declared to be narrower than T, either implicitly or explicitly, is to be taken to be the expression of the vocabulary designer's intention that this be a legitimate value for this property.

## 5.3 Data Model

### 5.3.1 UML Profile

The data model proposed in this document is fully defined in all detail through a UML model. UML is a large language and we have consciously restricted ourselves to a subset of the possible modelling elements. We have also added a few modelling elements using the extension mechanisms UML provides through stereotypes, tags and predefined data types. This combination of restriction and extensions is referred to as a UML Profile. The details of our profile are described in Appendix B.

### 5.3.2 Characterisation data model

As described in section 3.7, the model allows one to characterise the results of experiments statistically using the StatisticalSummary class. This part of the model addresses similar problems for simulations as does the Characterisation Data Model for observations. We have not followed that model in detail, but have tried to incorporate its main ideas, giving a new interpretation to some of these[19]. We believe the best way to reconcile the two approaches is to see both as specialisations of a more abstract model defining statistical characterisations of data products. A proposal for such a "domain model for characterisation was given in [32].

### 5.3.3 UTYPE

Section 4.1 describes how we generate UTYPEs for the different elements in our data model. The rules we use to do so have been subsumed in a draft for a Note on UTYPE-s by [17]. One problem we have with that Note is that the concepts used in the grammar, and that are direct reflections of syntactic modelling elements in our UML profile, have not been defined. For models defined with different UML syntax the grammar does not help.

---

[19] This follows ideas presented in China 2007, see
http://www.ivoa.net/internal/IVOA/InterOpMay2007DataModel/CharacterisationInTheDomain.ppt

Our assumption has been that a UTYPE should allow one to uniquely identify a concept in a data model. We do not assume that our particular form to do so needs to be taken over. But, as we describe in 4.1, *if* one wants to simply derive a list of unique strings to be associated to concepts that play a role in data models designed with our UML profile, these rules may help. Clearly if the syntax were to change we can accommodate that easily.

The effort on understanding what UTYPEs really are, how they are to be used, or defined is in our opinion not completed. But we feel that our approach is compatible with any possible interpretation, and sufficiently flexible to proposed changes in precise syntax, were they required.

# 6 References

## 6.1 Accompanying documents

[1] This document, at web address
http://ivoa.net/Documents/SimDM/20120503/index.html
[2] SimDM UML diagram obtained from MagicDraw :
http://ivoa.net/Documents/SimDM/20120503/uml/SimDM_DM.xml
[3] A PNG representation of the main diagram, 'all', in the model, extracted from MagicDraw in
http://ivoa.net/Documents/SimDM/20120503/uml/SimDM_DM.png
[4] "Intermediate representation" of the model. An XML document containing all relevant information from the model in a more readable format than XMI. This document is generated from the XMI and is itself the source of all other generated products.
http://ivoa.net/Documents/SimDM/20120503/uml/SimDM_INTERMEDIATE.xml[20]
[5] HTML representation of the SimDM in
http://ivoa.net/Documents/SimDM/20120503/html/SimDM.html
[6] XML schema documents derived from the data model and defining the representation of data model instances in XML. Divided over various documents. The "element schema" document defining all root elements can be found here: http://ivoa.net/Documents/SimDM/20120503/xsd/SimDM_root.xsd . All type schemas can be found in the same folder, http://ivoa.net/Documents/SimDM/20120503/xsd and sub-folders of it.
[7] Franck Le Petit et al., *Implementation of the Simulation Data Model*
http://www.ivoa.net/Documents/Notes/ImplementationSimDM/index.html

---

[20] The XML schema file defining the structure of the representation in [4] can be found there:
http://ivoa.net/Documents/SimDM/20120503/xsd/intermediateModel.xsd

## 6.2 Relevant IVOA documents

## 6.3 Other sources

[26] http://cds.u-strasbg.fr/twikiDCA/pub/EuroVODCA/Deliverables/EuroVO-DCA_D11_MPG_Final.pdf
[27] Martin Fowler (1997) *Analysis Patterns*
    Addison Wesley Longman, Inc
[28] Terry Halpin (2001) *Information Modeling and Relational Databases: From Conceptual Analysis to Logical Design*
    Morgan Kauffmann Publishers
[29] XML schema, http://www.w3.org/XML/Schema
[30] *MOF 2.0/XMI Mapping, V2.1.1*
    http://www.omg.org/spec/XMI/2.1/PDF
[31] *OMG Unified Modeling Language (OMG UML), Infrastructure* Version 2.2
    http://www.omg.org/spec/UML/2.2/Infrastructure/PDF/
[32] Gerard Lemson (2007) *Characterisation in the domain*
    Presentation at IVOA interoperability meeting Bejing 2007.
    http://www.ivoa.net/internal/IVOA/InterOpMay2007DataModel/CharacterisationInTheDomain.ppt
[33] http://en.wikipedia.org/wiki/Conceptual_data_model
[34] http://en.wikipedia.org/wiki/Logical_data_model
[35] http://en.wikipedia.org/wiki/Physical_data_model
[36] Jim Gray et al (2002) *Data Mining the SDSS SkyServer Database*
    http://www.sdss.jhu.edu/ScienceArchive/pubs/msr-tr-2002-01.pdf
[37] Gerard Lemson (2004) *Model Based Schema*
    PPT presented during Registry video conference 2004-05-13
    http://www.g-vo.org/www/uploads/Documentation/Registry_XSD_videocon20040513-14.ppt
[38] S. Bradner, RFC 2119: Key words for use in RFCs to Indicate Requirement Levels.
    http://www.rfc-editor.org/rfc/rfc2119.txt, 1997. IETF Request For Comments.
40

# Appendix A     History

Numerical computer simulations form an increasingly important component of astrophysical research. Such simulations are used to model astrophysical processes whose complexity precludes an analytical treatment. The subject of these simulations includes every possible astrophysical phenomenon, from the structure of stellar atmospheres, the formation of solar systems, the structure of galaxies and the description of their constituents, to the formation of the largest structures in the universe.

The simulations often result in predictions that can be compared to observations, but in general are much richer, including "observables" that can only be derived by indirect means from observations. These results can be very large, rivalling and often exceeding in size the largest observational catalogues. But they can also be relatively small, consisting of individual spectra of say a white dwarf, though often in collections resulting from parameter studies.

The design and execution of these simulations has become a specialised field of astrophysics, and is these days often performed in large collaborations. And while it is still true that their results are studied by these groups only, more and more of these theoretical data are being published online (see for instance the Appendix B of [26]).

Apart from limited support for publishing theoretical spectra in SSAP, there is as yet no IVOA standard dealing with the publication of simulations and their results. In earlier documents we have described the issues for defining such standards compared to the arguably simpler case of observational data sets (see for example [21] and [26]).

The proposal for a standard way of publishing simulations was formulated during a workshop in Cambridge, February 2006. The original idea was to create an analogue of the simple image access protocol (SIAP, [18]) for N-Body simulations: SNAP, the *Simple Numerical Access Protocol*. During the following interoperability meeting in Victoria, May 2006, the scope was expanded to include other types of simulation algorithms, and rephrased to something like "simulations that reproduce 3+1dimensional space time". It was felt furthermore that not only simulations themselves should be included, but also certain types of post-processing such as cluster finders, as long as their results are still aimed at producing a description of 3D space at one or more points in time. Over time requests have come in to generalise this scope even more, basically to enable any type of astrophysical simulation to be handled.

An important change that was decided in Victoria 2006 was that instead of the SIA protocol, the newer simple spectral access protocol (SSAP, [19]) should be followed as an example. This protocol's main difference with respect to SIAP was the explicit data model that was created for spectra and was used as motivation for the queryData metadata and the getData data format. Hence SNAP from the beginning had a double focus on a data model plus related query protocol on the one hand, and a data access and delivery specification on the other hand.
Shortly before the Trieste interop in the spring of 2008, it was decided to split SNAP up along these lines in two separate specifications: a specification for a *Simulation Database* (SimDB) which would support searching for interesting simulations and services providing access to them, and a *Simulation Data Access Protocol* (SimDAP) providing a specification for accessing simulation results.

SimDB on its own is still a rather complex specification. It has overlap with the efforts and results of many working groups, Data Model (DM), Registry, Data Access Layer (DAL), Semantics as well as being an integral part of the Theory Interest Group (TIG). This issue has been discussed in the Baltimore and Strasbourg interops, as it causes a potential problem for the standardisation process: an interest group cannot promote a document to a standard, but which a working group (WG) could do so. It was decided in Baltimore to postpone that decision by creating a focus group led by the TIG and with participation form the various WGs.

The current document is the result of a split in original Note that was written for SimDB. Such a split was proposed to simplify the standardisation process and after some refactoring was performed mid-2009. This current document is the first of these and deals exclusively with the data model (SimDM) and consequently has a natural place in the DM WG. The second document deals with the use of the data model for defining the model for a relational database and its related TAP query implementation as well as a service interface for uploading simulation descriptions to this database. It is not yet clear whether it can find a place in a single WG.

A parallel effort has been the proposal for a simpler access standard for small scale simulation, the Simple Self-describing Service protocol (S3, [15]). This was a result of an investigation started in the Cambridge 2007 interoperability meeting whether "micro-physics" simulations as they are sometimes called require special attention. For some time this was covered by SSA, at least as far as theory spectra were concerned. S3 is actually a direct reworking of an older Theoretical Spectral Access Protocol [20].

There were questions in the TIG whether S3 might be incorporated in SimDB and/or SimDAP. In the interoperability meeting in Victoria 2010 the decision was made that indeed this should be possible. The SimDM was shown to be able to incorporate the metadata for S3-like services, and indeed proposes extensions of that. It was decided that the S3 protocol should be merged with the SimDAP protocol, which from then on will be known by the name Simulation Data Access Layer (SimDAL).

# Appendix B  UML Profile[21]

The Simulation Data Model uses UML as the language for its specification. This is in accordance with decisions of the IVOA data model working group. One advantage of UML is that it is implementation neutral. It is a graphical language, consisting of "boxes and lines" that is very suitable for whiteboard discussions but allows one to model the concepts and relations in a static data model. It is also rich enough to allow one to describe all important data elements and relations.

In fact, UML is almost too rich. It is easy to become overwhelmed by the large number of possible syntactic elements to choose from for modelling a particular structure. Luckily UML allows one to formally define a subset of its language where one restricts oneself to a subset of the syntactic elements. Such a subset is called a *UML Profile*. Apart from creating a more restricted language, a Profile also allows on to assign new meanings to existing elements by defining *stereotypes* with associated properties (*tag definitions*). It is also possible to predefine classes and data types (see below) that can be reused by the data modeller.

In our modelling effort we have defined an initial implementation of a UML profile as created by MagicDraw. The profile[22] is contained in the UML file containing the SimDM data model. Here we give a list of the main elements that we use and give a short motivation for their inclusion in the language. It is our opinion that the DM working group should be ultimately responsible for a profile such as this, as it gives the possibility of defining a domain specific language for all IVOA data modelling efforts, thus giving some uniformity to those disparate efforts.

## B.1   Element

All elements mentioned below are specialisations of UML Element.
**Stereotypes**
- **<<modelelement>>**: This stereotype can be assigned to any UML *Element* and is used to define the **utype** tag on.
  Tags:
    - **utype [string]**: this holds the actual UTYPE that points to the other modelling element that is represented here.

## B.2   Model

---

[21] This Appendix could eventually be replaced by a Note on UML profiles if the DM WG is interested in organising such an effort. As this does currently not exist we have kept it in here to make the specification as self-contained as possible.
[22] Available under http://vo-urp.googlecode.com/svn/trunk/uml/IVOA%20UML%20Profile%20v-3.xml

This is the root of the complete model, contains all packages, classes etc. Also contains any imported profile.

**Stereotypes**
- **<<model>>**
  If the designer wants to annotate the model with the tags in this stereotype (s)he must explicitly associate this stereotype to the Model.
  Tags:
    - **author**: Indicates the author(s) of the model.
    - **title**: provides a long title to the model. The name of the model is assumed to be short.
    - **subject**: 0..* list of subjects in the sense of the Registry's subject attribute.

## B.3    Package

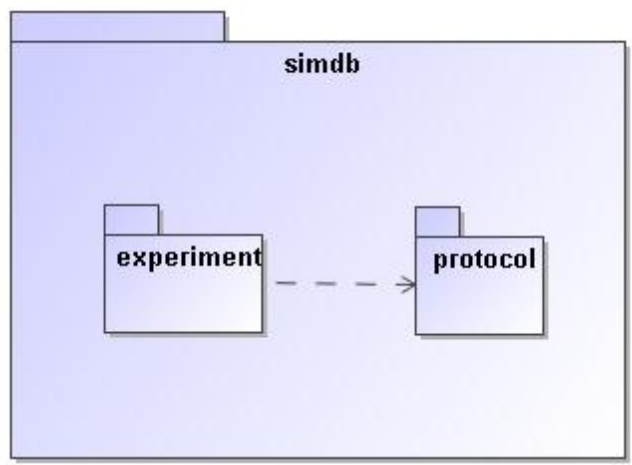

**Figure 17: This figure shows a package "simdb" that contains two other packages. Of these the experiment package depends on the protocol packages, which is indicated by the dashed arrow. See Figure 2 for the somewhat more complex package structure used in SimDM.**

A package groups related elements such as class definitions and possibly sub packages. Packages can depend on each other (indicated by the dashed line), which means that elements in one package can use elements in the target package in their definition. This relation is transitive. A package is similar to an XML namespace and in fact we map UML packages to XML namespaces in the XML schema mapping for the model described in 4.2.

## B.4    Class

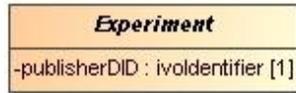

**Figure 18: A Class is a rectangular box, with the name of the class in boldface.**

Classes are the fundamental building blocks of a data model. A Class represents a full-fledged concept and is built up from properties and relations to other Classes. An important feature of Classes as opposed to DataTypes (see below) is that instances of Classes, i.e. objects, have their own, explicit identity[23]. That is, we want to assign an explicit identifier to each particular usage of this concept, for instance here to distinguish between various Experiment instances.

**Properties**:
- **isAbstract**
  Indicated by *italicised* name of the object. Implies that no instances can be made of the class, only of concrete (=non abstract) sub classes.

## B.5    *ValueType*

A ValueType represents a simple concept that is used to describe/define more complex concepts such as Classes. ValueType-s are, in contrast to Classes not separately identified. They are identified by their value. For example an integer is a value type; all instances of the integer value 3 represent the same integer.
In this profile ValueType-s are only represented using specialised examples. Attributes (see below) must have a ValueType as their datatype.

## B.6    *PrimitiveType*

PrimitiveTypes are the simplest examples of ValueTypes. They are represented by a single value only. A set of PrimitiveTypes is predefined in the IVOA profile (see Figure 19).

---

[23] This is admittedly a somewhat theoretical object-oriented concept.

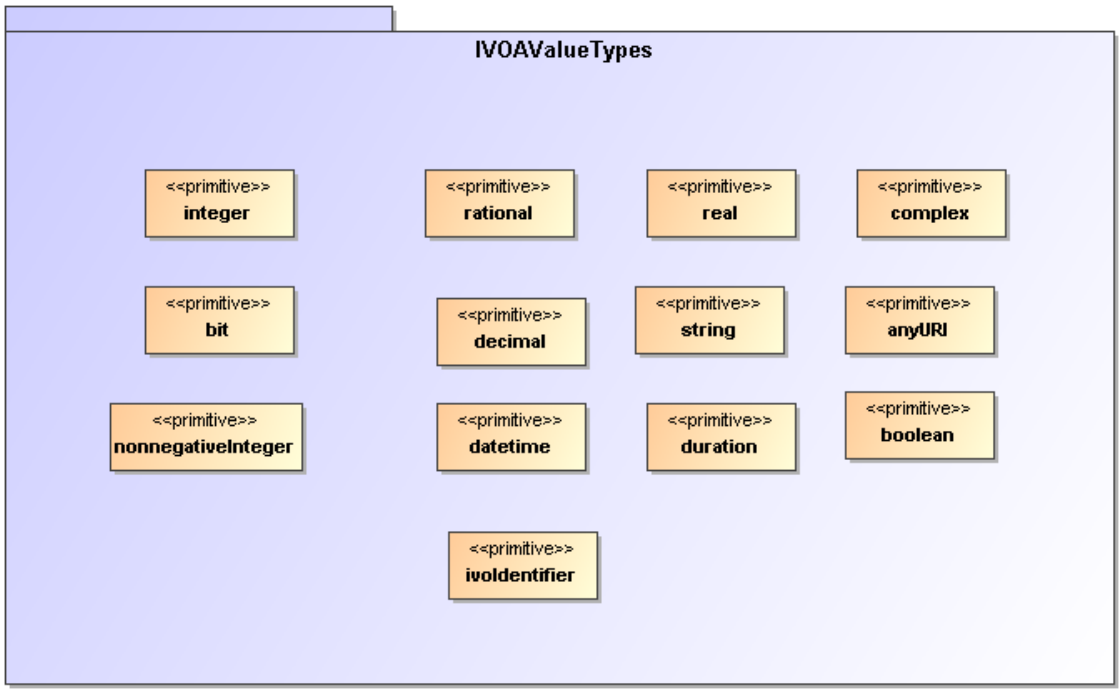

**Figure 19: The PrimitiveTypes that are predefined in the IVOA profile.**

## B.7    DataType

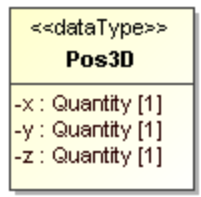

**Figure 20: Example of a structured datatype:Pos3D represents a position in 3D space and is defined using x, y and z attributes. The DataType symbol is distinguished from the Class by the <<dataType>> stereotype.**

A DataType is a ValueType that has more structure than a single value. This structure is modelled using Attributes, just as on ObjectTypes.

## B.8    Enumeration

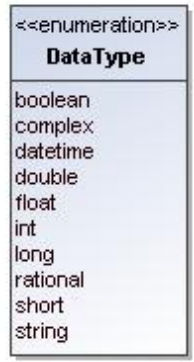

**Figure 21: An enumeration is indicated by a box with the name of the enumeration and the list of valid literals.**

An Enumeration is a ValueType that is defined by a list of valid values. These are the only values that instances of this data type can assume.

## *B.9    Attribute*

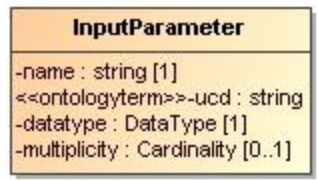

**Figure 22: An attribute is indicated by a line with a name, a datatype, an indication of the multiplicity and possibly a stereotype.**

An Attribute is a Property of a type (object type as well as structured data type). An attribute's data type is always a Value type, not an object type. For object type properties one should use a [Reference](#).

**Properties**
- data type
- multiplicity/cardinality: indicates the cardinality of the attribute (assumed to be 0..1, or 1. This is a relational bias based on normal form and the assumption that most databases do not allow storage of arrays in single columns.)
- 

**Stereotypes**
- <<attribute>>
  To assign further properties such as the tags this stereotype attribute must be explicitly assigned.
  Tag definitions
    - length [integer]: Constraint indicating that an attribute must have a specific fixed length. Is relevant only for attributes of type string.

- o maxLength [integer]: Constraint indicating that an attribute may at most have the indicated length. Is relevant only for attributes of type string. Is used in mappings to TAP to indicate the length of the corresponding column. Thist would seem to be very much an application specific feature and therefore belong to logical modelling. But this profile can be used for that purpose, hence it is included.
- o uniqueGlobally: Constraint indicating that only one instance of the type of the Class owning this attribute can have a given value. Globally should be read to mean globally in a given instance of the model, i.e. a database for example that stores instances of the model.
- o uniqueInCollection [boolean]: If true, indicates that the value of the attribute cannot be shared by the same attribute of any other instance of the Class owning this attribute that is in the same collection, i.e. has the same container object. In SimDB/DM an example is given by the name attribute of the InputParameter class for a given Protocol.
- <<ontologyterm>>
  There are many instances in the data model where we need to describe elements of the SimDB/Resource-s explicitly, because we do not have implicit information based on the context. Examples are the various properties of object types, the target objects and processes etc. Apart from a name and a description we then frequently add an attribute which is supposed to "label" the element according to an assumed standard list of terms.
  We model this using the <<ontologyterm>> stereotype. Attributes with this stereotype are assumed to take their values form such a predefined "ontology"[24].
  Tag definitions:
  - o ontologyURI
    A URL locating a standard (RDF|SKOS|OWL|???) document containing a list of terms from which the value for this attribute may be obtained. It is our opinion that the Semantics working group should be responsible for the definition of relevant ontologies (or semantic vocabularies, or thesauri, or ...) required for a given application domain, though the contents should be decided in cooperation with domain experts.

## B.10 Inheritance

---

[24] Possibly this should be a vocabulary, that at least is intended, and the stereotype might have to be called <<skosterm.., with tag definition named skosVocabulary.

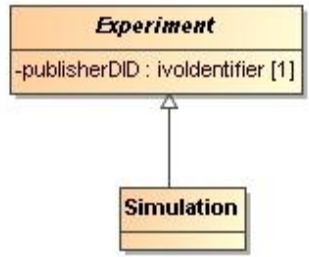

**Figure 23: Inheritance is indicated by a line with an open arrow from a subclass to its base class.**

Indicates the typical "is a" relation between the sub-class and its base-class (the one pointed at). In this profile we do not support multiple inheritances.

## B.11 Collection

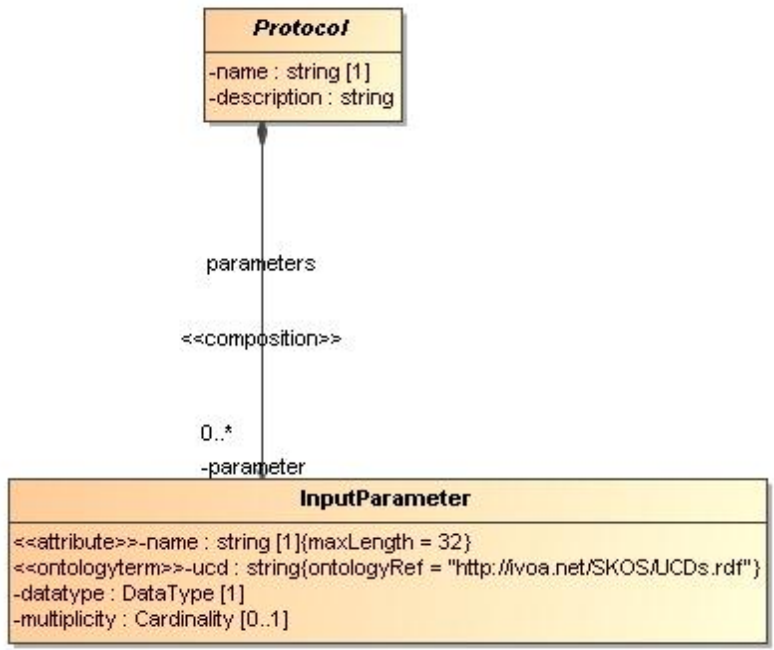

**Figure 24: The line with the filled diamond on one end and an arrow on the other indicates a (parent-child) composition relation, or collection, between the parent, on the side of the diamond; and the child, on the other side.**

This relation indicates a composition relation between one parent, container object and 0 or more child objects. The life cycles of the child objects are governed by that of the parent.
In UML a composition relation is represented by a binary association end.

## B.12 Reference

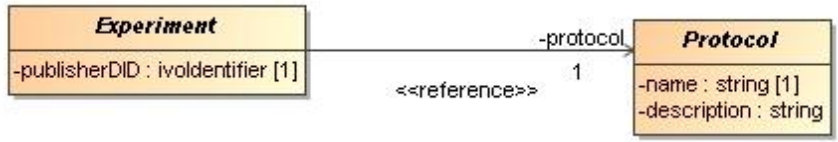

**Figure 25: A Reference is represented by a line connecting a class with another, referenced, class with an arrow on the referenced class. Note, the <<reference>> stereotype indication is not required.**

This is a relation that indicates a kind of usage, or dependency of one object on another. It is in general shared, i.e. many objects may reference a single other object. Accordingly the referenced object is independent of the "referee". In our profile the cardinality cannot be > 1.
For implementing the Reference in UML we use a shared, navigable binary association end.

## B.13     Subsets

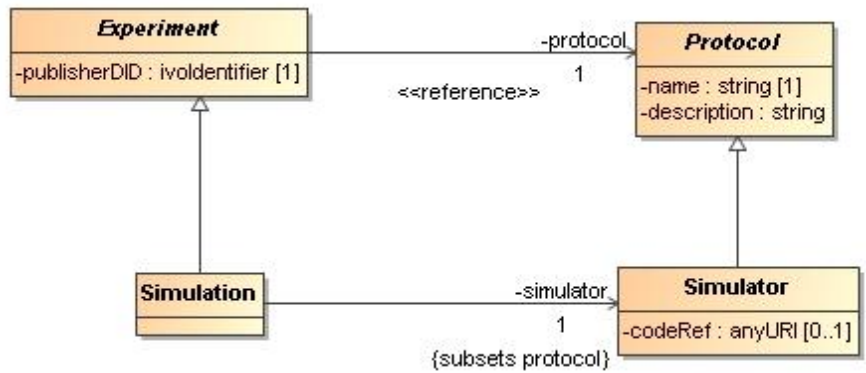

**Figure 26: The subsets property can only be assigned to a relation between two objects that are both subclasses of Classes that have an equivalent relation. It is indicated by the {subsets ...} annotation to the relevant association end.**

The "subsets" property, when associated to a Reference or Collection (in UML to the corresponding association end), indicates that a relation overrides the definition of a relation of the same name defined on a base class. It does so by specifying that the target class at the end point of the relation should be a subclass of the target class at the endpoint of the original, sub-setted relation.

## B.14     Instance diagrams

In the text we provide examples of instances of the UML model using instance diagrams. An example is the one shown in Figure 27.

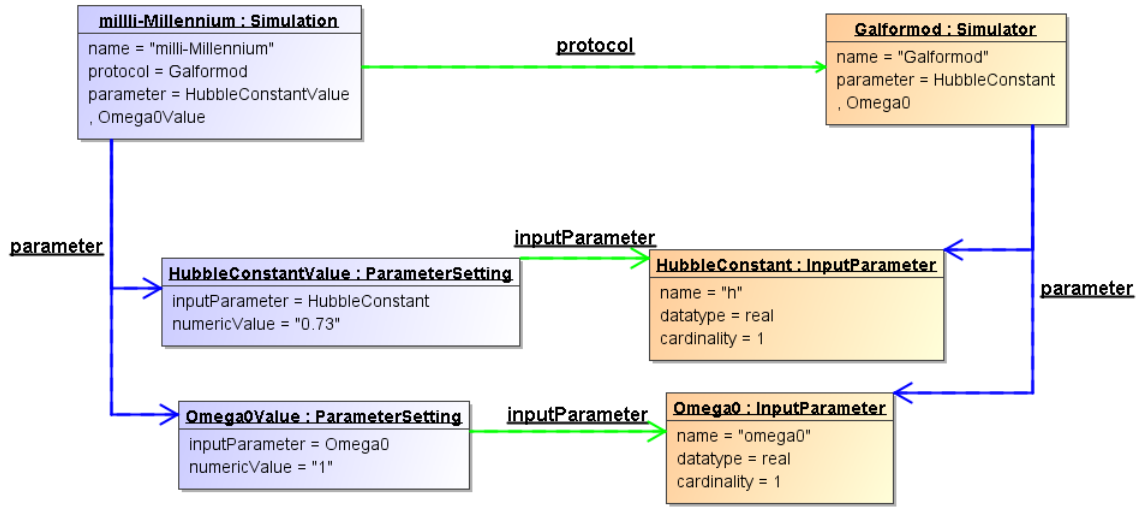

**Figure 27 Instance diagram example.**

Boxes represent objects, i.e. instance of a particular class. The name of the class is indicated in the header after the colon. The name before the colon is simply an identifier of the instance. The lower part of the box shows properties and their values. This includes both attributes and collection and reference relations. The latter are also represented using directed links between the related objects. Blue links represent composition relations, green lines reference relations.
The diagram therefore represents the following situation:
A Simulator named "Galformod" is defined with 2 parameters, "h" and "omega0". These both have data type "real" and cardinality "1". A Simulation was run using this Simulator as indicated by the "protocol" reference between the two objects. The name of the simulation is "milli-Millennium". It assigned values to the parameters, 0.73 to "h" and 1 to "omega0".
More complex instance diagrams are shown in various places in the text. All are snippets of the full example which is summarised in the next Appendix.

# Appendix C  Issues

We identify and discuss here several issues with the SimDM that need to be discussed further.

## C.1  Normalisation

The current version of the SimDB/DM is rather more *normalised* [9] than most of the other data models in the IVOA. We explain this concept based on a particular choice we made during the modelling process, and then we discuss the consequences of particular choices.

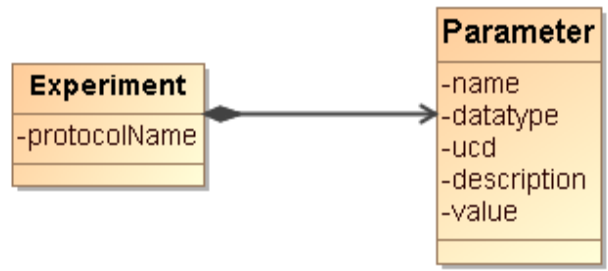

**Figure 28: Non-normalised model for experiments and parameters.**

At an earlier stage, the model was less normalised in the design of the input parameters of an experiment, as is illustrated in Figure 28. There was no separate protocol class, only an attribute *protocolName* on the **Experiment** class indicated the protocol by which the experiment was run. Also, the input parameters on the experiment were completely contained in a collection of **Parameter**-s. The **Parameter** class contained all the details, including *name* of the parameter, *description*, *ucd* etc. It also contained the *value* of the parameter in the experiment.

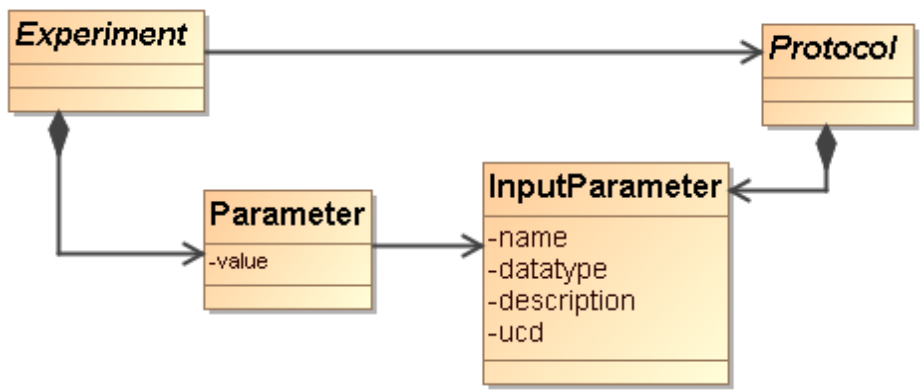

**Figure 29: Normalised modelling of experiments, protocols and their parameters.**

Currently the model treats parameter definitions and settings as in Figure 29. In this normalised design, the *Protocol* is given a class of its own, and it contains the input parameter collection. The *InputParameter* class does *not* contain a *value*, only the definition of the parameter: *name*, *datatype*, *ucd*, *description*. The values assigned to parameters in a given experiment are captured with the *ParameterSetting* class, contained in a collection of *Experiment*.

The motivation for this change of model was that a SimDB instance will in general contain many simulations (experiments) run with the same simulator code (protocol). In the old model, each experiment has to define the collection of input parameters with all details. In the new design this only has to be defined once, on the appropriate protocol. This clearly is less redundant, which is one important design goal of a normalised modelling approach. At the same time it provides an explicit identity to input parameters, which allows us to ask explicit questions about all parameter settings for a given, identified parameter. In the old model this is only indirectly possible, using equality of the name of input parameters for all experiments having the same *protocolName*. Now we can ask for all experiments with the same protocol reference, and look for parameter settings with the same input parameter reference. This is arguably a more "correct" model of reality.

There are therefore advantages to normalisation, but there are also disadvantages. We need to realise these and make choices that optimise the usability of the data model. One of the main disadvantages is that *references*, which naturally have to be introduced when normalising a model, are more difficult to deal with than most of the other modelling elements, particularly in some physical representations (see below). When defining a new experiment, one will have to find the input parameter that one needs to set, and instead of simply giving name/value, one needs to represent the reference to the parameter. For this one may have to extract the protocol as stored in the SimDB and find the appropriate identifiers of the input parameters. In this sense an *Experiment* definition becomes less self-contained. It depends on the details of the *registered Protocol*. This protocol is registered separately and necessarily at an earlier point in time.

This puts strong requirements on SimDB implementations to maintain referential integrity, something which will be even harder to achieve if we were to allow cross-SimDB referencing. In one advanced usage scenario the UC San Diego version of SimDB registers the Enzo[25] simulator, whilst the Italian SimDB allows registration of simulations that used it and reference the remote protocol[26].

Similarly a query language needs to be able to handle with this level of indirection. For example in a relational database one needs to write joins between ParameterSetting and InputParameter. For expert SQL users this is not a problem, but is something to get used to. For simpler query languages, those not allowing joins, like TAP/Param, asking meaningful queries becomes very difficult. One way around this problem could be to add some view definitions to the model. In relational databases, views are predefined, named SQL queries that can be treated as if they were tables when querying the database. It is quite straightforward to define some SQL queries that as it were denormalise the model and put the input parameter definition back under the experiment together with the value. This way one may protect users of the database from the high level of normalisation.

---

[25] http://lca.ucsd.edu/portal/software/enzo
[26] We actually do not support this scenario in the current version of SimDB.

## C.2    Quantities and Units

At various locations in the SimDM data model numerical values can be defined, for example in parameter settings or the characterisation of properties of representation object type collections. Often these numerical values will need to have a unit. The IVOA has two ways of dealing with units. Either units are fixed explicitly for properties/parameters in protocol or data model, sometimes depending on the small list of possible UCDs. Alternatively units are explicitly stated, for example in VOTable. At the moment we support the second mode, especially because, as is true for VOTable, we do not know what kind of property is being used. To this end we introduce a value type in the model, `Quantity`, which contains a value and a unit, and which is the data type of various value attributes, for example in *NumericalParameterSetting* or *Characterisation*. In the XML schema this is translated to a `complexType` with 2 elements, in the relational database schema to two columns, one with the value, and one with the unit. It is in the use of the relational schema that we anticipate problems with this approach, especially in the query protocol to SimDB. Consider the typical science question: return all N-body simulations with particle mass roughly $10^{10}$ $M_{sun}$. In SimDB this would be need to be translated in an ADQL query which contains the unit column explicitly. Allowing users' freedom of registering SimDB resources using any units they desire can lead to resource containing, for the same observable "N-body particle mass", values with a whole range of units. To provide reasonable support to users requires the SimDB implementation to be able to do the automated transformation. We propose in SimDB to use ADQL/TAP as the query interface. If units are stored explicitly users can phrase queries using these.

An alternative approach is to mandate stating values for properties with a given UCD (or other semantic label) always with the same units. This would solve the query problem but poses others. For one it may be very (too?) unnatural for users to be forced to use meters for cosmological simulations, or megaparsec for simulations in the solar system. Related to this is the probably contentious discussion of what units to assign to what UCD. One might choose SI or cgs units, but these are not always very useful or natural.

## C.3    Linking services, experiments and other resources.

Compared to the Domain Model discussed in 2.2, the Simulation Data Model proposed here lacks concepts representing the *storage* of results. The reason is that whereas we can define the concept of a Result as collections of Objects quite satisfactory, we have shied away of trying to model the precise way these results are actually stored in files or a database. There are simply too many possible and actual ways in which results can be stored in a file system. In general, a result, or snapshot in our model, cannot be modelled with a simple reference to a file or table in which it is stored. Large results may be split up over many files, stored as structures, or in arrays. We have therefore decided not to open this can of worms (or reopen it, see the Quantity data model [http://www.ivoa.net/internal/IVOA/IvoaDataModel/qty23.pdf](http://www.ivoa.net/internal/IVOA/IvoaDataModel/qty23.pdf)).

Instead we assume the existence of web services that allow users access to the results of SimDB experiments. Some of these services may implement a standard protocol as defined by SimDAL, or they may be custom services. The precise way to relate experiments to services and what can be inferred about how to call them is the task of the SimDAL protocol.

In the model actually web services are related to resources in general. This can be used to represent services that gives access to a set of experimental results from some project, or that can for example visualise any result of experiments performed according to a fixed protocol, for example generic Gadget-format visualisers.